\documentclass[twocolumn,aps,prb,notitlepage,nofootinbib,longbibliography]{revtex4-2}
\usepackage[latin9]{inputenc}
\usepackage{verbatim}
\usepackage{bm}
\usepackage{amsmath}
\usepackage{amssymb}
\usepackage{graphicx}
\usepackage{amsfonts}

\begin{document}
\title{Suppression of superconducting parameters by correlated quasi-two-dimensional magnetic fluctuations}
\author{A. E. Koshelev}
\affiliation{Materials Science Division, Argonne National Laboratory, Argonne,
Illinois 60439}
\date{\today }
\begin{abstract}
We consider a clean layered magnetic superconductor in which a continuous
magnetic transition takes place inside a superconducting state. We assume
that the exchange interaction between superconducting and magnetic
subsystems is weak so that superconductivity is not destroyed at the
magnetic transition. A representative example of such material is
RbEuFe$_{4}$As$_{4}$. We investigate the suppression of the superconducting
gap and superfluid density by correlated magnetic fluctuations in
the vicinity of the magnetic transition. %
The influence of nonuniform exchange field on superconducting parameters
is very sensitive to the relation between the magnetic correlation
length, $\xi_{h}$, and superconducting coherence length $\xi_{s}$
defining the 'scattering' ($\xi_{h}<\xi_{s}$) and 'smooth' ($\xi_{h}>\xi_{s}$) regimes. As a small uniform exchange field does not affect the superconducting gap and superfluid
density at zero temperature, smoothening of the spatial variations
of the exchange field reduces its effects on these parameters. We
develop a quantitative description of this 'scattering-to-smooth' crossover 
for the case of quasi-two-dimensional magnetic fluctuations
realized in RbEuFe$_{4}$As$_{4}$. 
Since the magnetic-scattering  energy scale 
is comparable with the gap in the crossover region, the standard quasiclassical approximation is not applicable  and full microscopic treatment is required. 
We find that the corrections to
both the gap and superfluid density increase proportionally to 
$\xi_{h}$ until it remains much smaller than $\xi_{s}$. In the opposite
limit, when the correlation length exceeds the coherence length both
parameters have much weaker dependence on $\xi_{h}$. Moreover, the
gap correction may decrease with increasing of $\xi_{h}$ in the immediate
vicinity of the magnetic transition if it is located at temperature
much lower than the superconducting transition. We also find that
the crossover between the two regimes is unexpectedly broad: the standard
scattering approximation becomes sufficient only when $\xi_{h}$ is
substantially smaller than $\xi_{s}$.
\end{abstract}
\maketitle

\section{Introduction\label{sec:Introduction}}

\begin{figure*}
\includegraphics[height=2.1in]{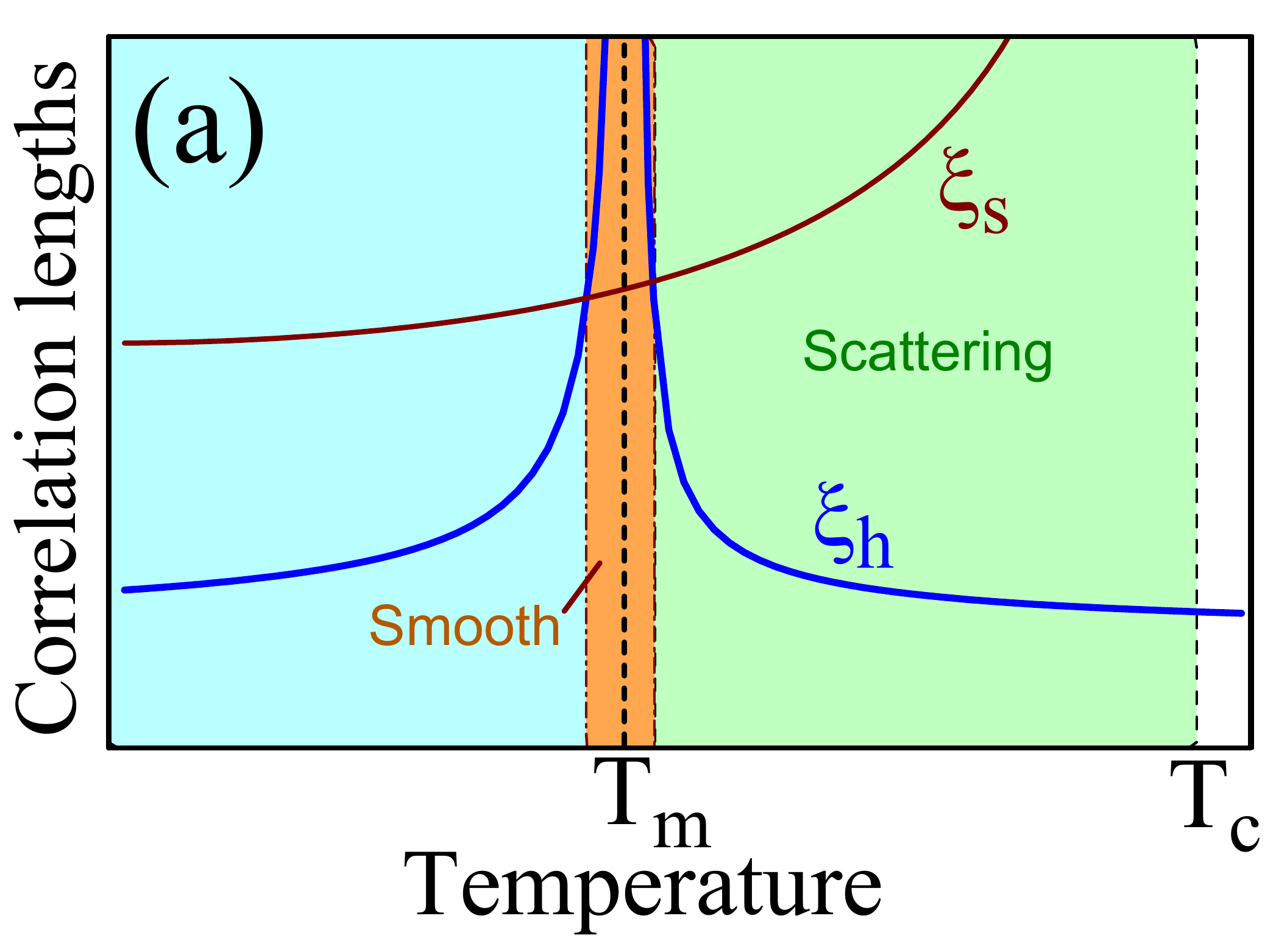}\includegraphics[height=2.1in]{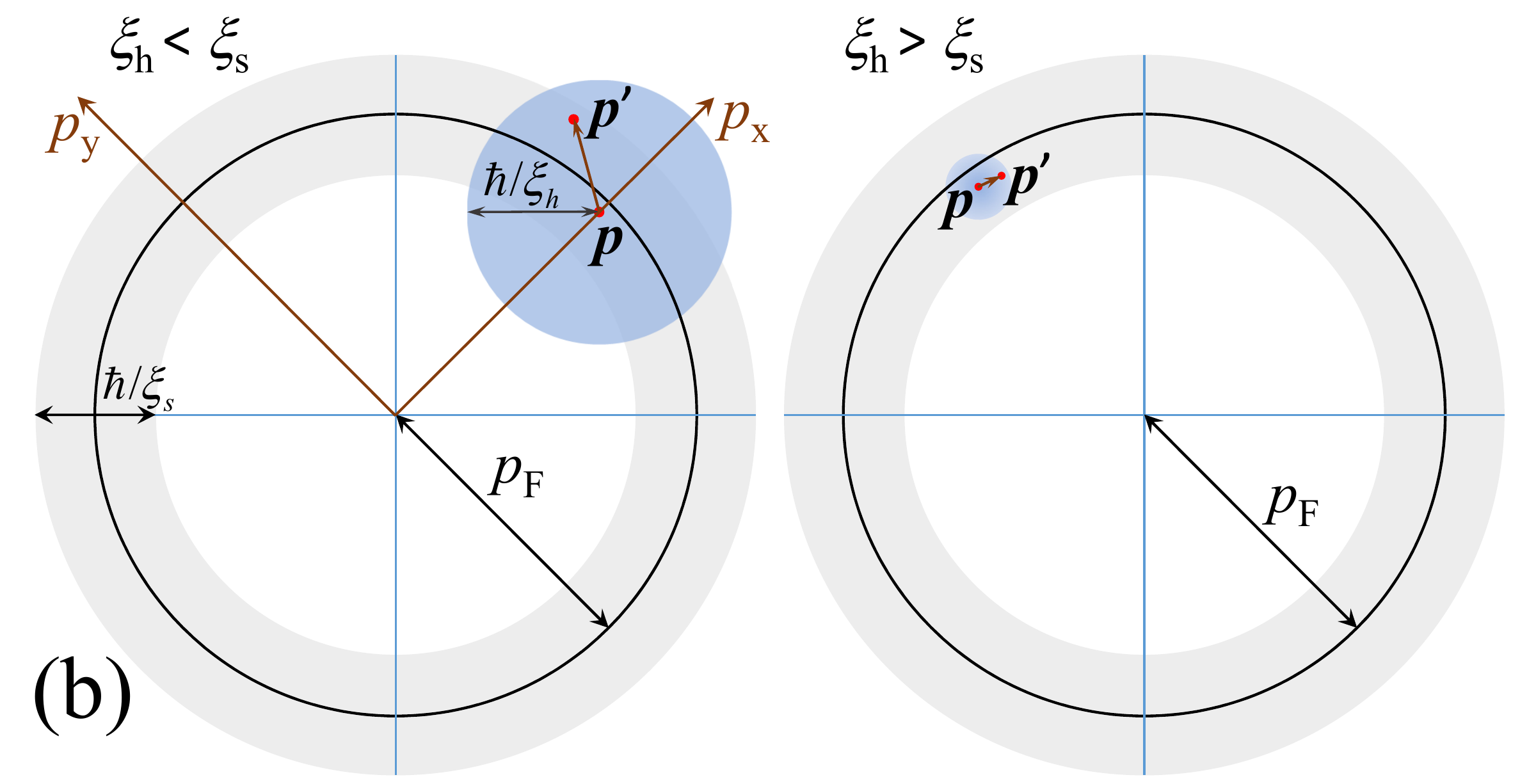}\caption{(a)Schematic temperature dependences of the magnetic correlation length
$\xi_{h}$ and superconducting coherence length $\xi_{s}$. The influence
of the fluctuating magnetic moments on superconductivity is very different
in the regions $\xi_{h}>\xi_{s}$ and $\xi_{h}<\xi_{s}$. (b)Typical
scales in momentum space characterizing scattering on magnetic fluctuations
for two relations between  $\xi_{h}$
and  $\xi_{s}$ in the case $\xi_{h}\gg k_{F}^{-1}$
with $\hbar k_{F}=p_{F}$. The small circle illustrates the small-angle
scattering on magnetic fluctuations with the range $\left|\boldsymbol{p}-\boldsymbol{p}^{\prime}\right|\sim\hbar/\xi_{h}$
and the ring with width $\hbar/\xi_{s}$ illustrate the range relevant
for superconductivity.}
\label{Fig:FS-MagnFluctLength}
\end{figure*}

Since the seminal work of Abrikosov and Gor\textquoteright kov (AG)\cite{AbrGorJETP61} 
and its extensions\cite{SkalskiPhysRev.136.A1500,MakiInParks1969,SchlottmannJLTP1975},
the pair breaking by magnetic scattering has been established as a
key concept in the physics of superconductivity. 
Its applications extend far beyond the original physical system for which the theory was developed, singlet superconductors with dilute magnetic impurities.
In particular, the magnetic pair-breaking scattering strongly influences properties of superconducting materials containing an embedded periodic lattice of magnetic rare-earth
ions.
Several classes of such materials are known at present including
magnetic Chevrel phases $\mathit{RE}$Mo$_{6}$\textit{X}$_{8}$ ($\mathit{RE}$=rare-earth
element and \textit{X}=S, Se), ternary rhodium borides $\mathit{RE}$Rh$_{4}$B$_{4}$\cite{BulaevskiiAdvPhys85,WolowiecPhysC15,KulicBuzdinSupercondBook,MapleFischerBook1982},
the rare-earth nickel borocarbides $\mathit{RE}$Ni$_{2}$B$_{2}$C\cite{MullerRoPP2001,GuptaAdvPhys06,MazumdarPhysC2015},
and recently discovered Eu-based iron pnictides\cite{Zapf2017,Liu2016a,Liu2016,KawashimaJPSJ2016,Bao2018}.
Some of these compounds experience a magnetic-ordering transition 
inside the superconducting state. Depending on the strength of the exchange
interaction between the rare-earth moments and conducting electrons, 
the magnetic transition may either destroy superconductivity or leave
it intact. In any case, in the paramagnetic state, the fluctuating
magnetic moments suppress superconductivity via magnetic scattering,
similar to magnetic impurities. Near the ferromagnetic transition,
the moments become strongly correlated which enhances the suppression.
The AG theory has been generalized to describe this enhancement in
several theoretical studies \cite{RainerZPhys1972,MachidaJLTP1979}.
A straightforward generalization, however, is only possible when the
magnetic correlation length $\xi_{h}$ is shorter than the superconducting
coherence length $\xi_{s}$ and this condition was always assumed
in all theoretical works. For a continuous magnetic transition, there
is always temperature range where this condition is violated, see
Fig.~\ref{Fig:FS-MagnFluctLength}(a). A small uniform exchange field
does not modify superconducting gap in clean materials at zero temperature
\cite{SarmaJPCS1963}, because, in absence of free quasiparticles,
the exchange field does not generate spin polarization of the Cooper-pair
condensate. This observation indicates that, once the exchange field
becomes smooth at the scale of coherence length, its efficiency in
suppressing superconducting parameters at low temperatures diminishes.
We can conclude that the existing treatments of the impact of correlated
magnetic fluctuations on superconductivity are incomplete. A full
theoretical description of this phenomenon requires consideration
of the crossover between the \textquoteleft scattering\textquoteright{} 
and \textquoteleft smooth\textquoteright{} regimes illustrated in
Fig.~\ref{Fig:FS-MagnFluctLength}(a). For most magnetic superconductors,
however, such full theory would be a mostly academic exercise, because
the coherence length is typically much larger than the separation
between magnetic ions. Consider, for example, the magnetic nickel
borocarbide ErNi$_{2}$B$_{2}$C, which has the superconducting transition
at T$_{c}\approx$11 K and magnetic transition at T$_{m}\approx$6
K \cite{MullerRoPP2001,GuptaAdvPhys06}. Its c-axis upper critical field has linear slope 0.3 T/K near T$_{c}$ \cite{BudkoPhysRevB.61.R14932},
from which we can estimate the in-plane coherence length at T$_{m}$
as $\xi_{s}(T_{m})\approx$18 nm which is $\sim54$ times larger than
the distance between the Er$^{3+}$ moments. Therefore, in this and
similar materials the magnetic correlation length exceeds the coherence
length only within an extremely narrow temperature range near the
magnetic transition. The situation is very different, however, in Eu-based layered
iron pnictides, such as RbEuFe$_{4}$As$_{4}$\cite{Liu2016a,KawashimaJPSJ2016,Bao2018,Smylie2018}.
The latter material has the superconducting transition at 36.5 K and
the magnetic transition at 15K. The magnetism is quasi-two-dimensional:
the Eu$^{2+}$ moments have strong ferromagnetic interactions inside
the magnetic layers with easy-plane anisotropy \cite{WillaPhysRevB.99.180502}
and weak interactions between the magnetic planes leading to helical
interlayer order\cite{IidaPhysRevB.100.014506,IslamPreprint2019}.
Due to the quasi-two-dimensional nature of magnetism, the in-plane
magnetic correlation length smoothly grows within an extended temperature
range. Another relevant material's property is a very short in-plane coherence
length, $\sim$1.5--2 nm, which is only 4--6 times
larger than the distance between the magnetic ions. As a consequence,
contrary to most magnetic superconductors, the magnetic correlation
length exceeds the coherence length within a noticeable temperature
range near the magnetic transition. Therefore, for the magnetic iron pnictides,
the crossover between the \textquoteleft scattering\textquoteright{}
and \textquoteleft smooth\textquoteright{} regimes is very relevant.
Recent vortex imaging in RbEuFe$_{4}$As$_{4}$ with scanning Hall-probe
spectroscopy revealed a significant increase of the London penetration
depth in the vicinity of the magnetic transition\cite{CollombPreprint2020}.
This suggests that the exchange interaction between Eu$^{2+}$ moments
and Cooper pairs leads to substantial suppression of superconducting
parameters near $T_{m}$.

The goal of this paper is to develop a quantitative theoretical description
of the influence of correlated magnetic fluctuations on the superconducting
gap and supercurrent response with a proper treatment of the crossover
at $\xi_{h}\sim\xi_{s}$. The problem occurs to be technically challenging
because in the crossover region the probability of magnetic scattering
varies at the energy scale comparable with the temperature or the
gap. This forbids the standard energy integration necessary for the quasiclassical
approximation and requires a full microscopic consideration. In this
consideration, one has to include the self-energy correction to the
electronic spectrum and maintain the energy dependence of the scattering
probability. 
As this accurate analysis is rather complicated, we utilize several simplifying assumptions.
We limit ourselves to the case of weak exchange interaction
and consider only the lowest-order corrections. We also assume the
static approximation for magnetic fluctuations. This assumption is
justified when typical frequency scale for magnetic fluctuations is
smaller than the superconducting gap. Due to the critical slowing down,
this always becomes valid sufficiently close to the transition.
In the scattering regime, the dynamic effects have been investigated
in several theoretical papers, see, e.g., \cite{RainerZPhys1972,CoffeyPhysRevB.27.2740,SchossmannDynJLTP1987}.
The behavior is also sensitive to the dimensionality of magnetic fluctuations.
Having in mind application to layered magnetic superconductors, such
as RbEuFe$_{4}$As$_{4}$, we assume quasi-two-dimensional magnetic
fluctuations. In this case the discussed effects are more pronounced
than for three-dimensional magnetic fluctuations\cite{MachidaJLTP1979}.

The paper is organized as follows. In Sec.~\ref{sec:Model}, we introduce
the model for layered magnetic superconductors. In Sec.~\ref{sec:Scattering},
we evaluate the self energy caused by scattering by correlated magnetic
fluctuations for arbitrary relation between the magnetic correlation
length and coherence length and develop a quantitative description
of the crossover between the scattering and smooth regimes. In Sec.~\ref{sec:Gap}, we use these results to evaluate the exchange correction to the
gap. In Sec.~\ref{sec:Kernel}, we evaluate the leading correction
to the electromagnetic kernel accounting for the vertex correction.
Also, in Appendix \ref{app:Magn-Scatt-Lond} this correction is evaluated
in the scattering regime with quasiclassical approach. Finally, in Sec.~\ref{sec:Discussion},
we discuss the results and illustrate them by plotting representative temperature dependences for the parameters roughly corresponding to RbEuFe$_{4}$As$_{4}$.

\section{Model\label{sec:Model}}

We consider a layered material composed of superconducting and magnetic
layers described by the Hamiltonian
\begin{equation}
\mathcal{H}=\hat{\mathcal{H}}_{\mathrm{S}}+\hat{\mathcal{H}}_{\mathrm{M}}+\hat{\mathcal{H}}_{\mathrm{MS}},\label{eq:SC-Magn-Hamilt}
\end{equation}
where 
\begin{align}
\hat{\mathcal{H}}_{\mathrm{S}} & =\sum_{n,\mathbf{p}_{\parallel},\sigma}\xi_{\mathrm{2D}}(\mathbf{p}_{\parallel})a_{n,\sigma}^{\dagger}(\mathbf{p}_{\parallel})a_{n,\sigma}(\mathbf{p}_{\parallel})\nonumber \\
+ & \!\sum_{n,\mathbf{p}_{\parallel},\sigma}\!t_{\bot}\left[a_{n+1,\sigma}^{\dagger}(\mathbf{p}_{\parallel})a_{n,\sigma}(\mathbf{p}_{\parallel})\!+\!a_{n-1,\sigma}^{\dagger}(\mathbf{p}_{\parallel})a_{n,\sigma}(\mathbf{p}_{\parallel})\right]\nonumber \\
- & \sum_{n,\mathbf{p}_{\parallel}}\left[\Delta a_{n,\uparrow}^{\dagger}(\mathbf{p}_{\parallel})a_{n,\downarrow}^{\dagger}(-\mathbf{p}_{\parallel})\!+\!\Delta^{\ast}a_{n,\downarrow}(-\mathbf{p}_{\parallel})a_{n,\uparrow}(\mathbf{p}_{\parallel})\right]\label{eq:HS}
\end{align}
is the standard BCS Hamiltonian describing a layered superconductor.
Here $\sigma$ is spin index, $\xi_{\mathrm{2D}}(\mathbf{p}_{\parallel})=\varepsilon_{\mathrm{2D}}(\mathbf{p}_{\parallel})-\mu$
is the single-layer spectrum, $t_{\bot}$ is the interlayer hopping
integral, and $\Delta$ is the superconducting gap. The full 3D spectrum
for this model is $\xi_{\mathbf{p}}\!=\!\xi_{\mathrm{2D}}(\mathbf{p}_{\parallel})\!+\!2t_{\bot}\cos\left(p_{z}s\right)$.
However, its exact shape has a very little effect on further consideration.
The second term, $\hat{\mathcal{H}}_{\mathrm{M}}$, describes the quasi-two-dimensional
magnetic subsystem leading to a continuous phase transition at $T_{m}$.
The last term 
\begin{equation}
\hat{\mathcal{H}}_{\mathrm{MS}}  \!=\!\!\sum_{n,m,\mathbf{R}}\!
\int\!\!d^{2}\boldsymbol{r} J_{nm}\!\left(\mathbf{r}\!-\!\mathbf{R}\right)
\mathbf{S}_{m}(\mathbf{R})\hat{\mathbf{\boldsymbol{\sigma}}}_{\alpha\beta}a_{n,\alpha}^{\dagger}(\mathbf{r})a_{n,\beta}(\mathbf{r})\label{eq:HSM}
\end{equation}
describes the interaction between the magnetic and superconducting
layers with the strength set by the nonlocal exchange
constants $J_{nm}\left(\mathbf{r}\!-\!\mathbf{R}\right)$. Here the index $m$ marks magnetic layers, $\hat{\mathbf{\boldsymbol{\sigma}}}$ is Pauli-matrix vector, and summation is assumed over the spin indices $\alpha$ and $\beta$.
We can
rewrite the interaction term as 
\begin{equation}
\hat{\mathcal{H}}_{\mathrm{MS}}\!=\!-\!\sum_{n}\int d^{2}\boldsymbol{r}a_{n,\alpha}^{\dagger}(\boldsymbol{r})\boldsymbol{h}_{n}(\boldsymbol{r})\hat{\boldsymbol{\sigma}}_{\alpha\beta}a_{n,\beta}(\boldsymbol{r}),\label{eq:HSMh}
\end{equation}
where 
\begin{equation}
\boldsymbol{h}_{n}(\boldsymbol{r})\!=\!-\!\sum_{m,\boldsymbol{R}}J_{nm}(\boldsymbol{r}\!-\!\boldsymbol{R})\boldsymbol{S}_{m}(\boldsymbol{R})\label{eq:ExField}
\end{equation}
is the effective exchange field acting on spins of conducting electrons.
It can be split into the average part $\bar{\boldsymbol{h}}$ due
to either polarization of the moments by the magnetic field or spontaneous
magnetization in the ordered state and the fluctuating part $\tilde{\boldsymbol{h}}_{n}(\boldsymbol{r})$,
$\boldsymbol{h}_{n}(\boldsymbol{r})=\bar{\boldsymbol{h}}+\tilde{\boldsymbol{h}}_{n}(\boldsymbol{r})$,
\begin{subequations}
\begin{align}
\bar{\boldsymbol{h}} & =-\sum_{m,\boldsymbol{R}}J_{nm}(\boldsymbol{r}-\boldsymbol{R})\boldsymbol{\bar{S}},\label{eq:hav}\\
\tilde{\boldsymbol{h}}_{n}(\boldsymbol{r}) & =-\sum_{m,\boldsymbol{R}}J_{nm}(\boldsymbol{r}-\boldsymbol{R})\tilde{\boldsymbol{S}}_{m}(\boldsymbol{R}).\label{eq:hfl}
\end{align}
\end{subequations}The fluctuating part of the exchange field also
depends on time. We assume that the time scales of magnetic fluctuations
exceeds time scales relevant for superconductivity and employ the quasistatic
approximation. This assumption is justified near the transition due
to the critical slowing down. The fluctuating part is characterized
by the correlation function
\begin{align}
 & \left\langle \tilde{\boldsymbol{h}}_{n}(\boldsymbol{r})\tilde{\boldsymbol{h}}_{n'}(\boldsymbol{r}')\right\rangle \label{eq:h-correl}\\
\! & =\!\!\sum_{m,\boldsymbol{R},\boldsymbol{R}'}J_{nm}(\boldsymbol{r}\!-\!\boldsymbol{R})J_{n'm}(\boldsymbol{r}'\!-\!\boldsymbol{R}^{\prime})\left\langle \tilde{\boldsymbol{S}}_{m}(\boldsymbol{R})\tilde{\boldsymbol{S}}_{m}(\boldsymbol{R}^{\prime})\right\rangle .\nonumber
\end{align}
Here we neglected correlations between different magnetic layers.
In the following, we limit ourselves to the case when the uniform
field, $\bar{\boldsymbol{h}}$, can be neglected. This corresponds
to the paramagnetic state and ordered state near the transition in
the absence of an external magnetic field. We will also neglect correlations between different conducting layers and drop the layer index, $\left\langle \tilde{\boldsymbol{h}}_{n}(\boldsymbol{r})\tilde{\boldsymbol{h}}_{n'}(\boldsymbol{r}')\right\rangle \rightarrow \delta_{n,n^{\prime}}\left\langle \tilde{\boldsymbol{h}}(\boldsymbol{r})\tilde{\boldsymbol{h}}(\boldsymbol{r}')\right\rangle $.
This corresponds to the two-dimensional approximation for magnetic fluctuations.
The spin correlation function is related to the nonlocal spin susceptibility $\chi(\boldsymbol{r}-\boldsymbol{r}')$.
Sufficiently close to the magnetic transition, the spin correlation
length exceeds the range of $J_{nm}(\boldsymbol{r}-\boldsymbol{R})$
and we can approximate
\begin{equation}
\left\langle \tilde{\boldsymbol{h}}(\boldsymbol{r})\tilde{\boldsymbol{h}}(\boldsymbol{r}')\right\rangle \approx\sum_{m}\mathcal{J}_{nm}^{2}\left\langle \tilde{\boldsymbol{S}}_{m}(\boldsymbol{r})\tilde{\boldsymbol{S}}_{m}(\boldsymbol{r}')\right\rangle .\label{eq:hcorr-Scorr}
\end{equation}
with $\mathcal{J}_{nm}=\sum_{\boldsymbol{R}}J_{nm}(\boldsymbol{r}-\boldsymbol{R})$.
Away from the transition, however, the nonlocality of the exchange
interaction may have substantial influence on the amplitude and extent
of the exchange-field correlations. We neglect these complications and assume the simplest shape of the correlation function of $\tilde{\boldsymbol{h}}(\boldsymbol{r})$ defined by a single length scale,  the in-plane magnetic
correlation length $\xi_{h}$,%
\begin{align}
\left\langle \tilde{h}_{\alpha}(\boldsymbol{r})\tilde{h}_{\beta}(\boldsymbol{r}^{\prime})\right\rangle  & =\frac{h_{0}^{2}}{2}\delta_{\alpha\beta}f_{h}\left(\left|\boldsymbol{r}\!-\!\boldsymbol{r}^{\prime}\right|/\xi_{h}\right),\label{eq:FlExFCorr}
\end{align}
where $f_{h}(0)\!=\!1$, and the parameter $h_{0}^{2}\!=\!\left\langle \tilde{\boldsymbol{h}}^{2}\right\rangle \!\approx\!\sum_{m}\mathcal{J}_{nm}^{2}\left\langle \tilde{\boldsymbol{S}}^{2}\right\rangle $
weakly depends on temperature. The Fourier transform of the correlation
function is
\begin{align}
\left\langle \left|\tilde{\boldsymbol{h}}_{\mathbf{\mathbf{q}}}\right|^{2}\right\rangle  & \!=\!sh_{0}^{2}\int\!d^{2}\boldsymbol{r}f_{h}\left(\frac{r}{\xi_{h}}\right)\exp\left(i\boldsymbol{qr}\right)\!=\!sh_{0}^{2}\xi_{h}^{2}\tilde{f}_{h}\left(\xi_{h}q\right).\label{eq:FlExFl-Fourier}
\end{align}
Here we assume a conventional Lorentz shape for the $q$ dependence,  $\tilde{f}_{h}\left(\xi_{h}q\right)\!=\!C_{h}/\left(1\!+\!\xi_{h}^{2}q^{2}\right)$
with $C_{h}\!=\!2\pi\int_{0}^{\infty}\!f_{h}\left(x\right)xdx$. In
real space, this corresponds to
\begin{equation}
f_{h}\left(\frac{r}{\xi_{h}}\right)\!=\!\xi_{h}^{2}\!\int\!\frac{d^{2}\boldsymbol{q}}{(2\pi)^{2}}\frac{C_{h}}{1\!+\!\xi_{h}^{2}q^{2}}\exp(i\boldsymbol{qr})\!=\!\frac{C_{h}}{2\pi}K_{0}\left(\frac{r}{\xi_{h}}\right).\label{eq:fhr}
\end{equation}
The logarithmic divergency $K_{0}(r/\xi_{h})\!\propto\!\ln\left(\xi_{h}/r\right)$
has to be terminated at the distance between neighboring moments $r\!\sim\!a$
. Since the function $f_{h}\left(r/\xi_{h}\right)$ is normalized
by the condition $f_{h}(0)=1$, this means that $C_{h}\!\approx\!2\pi/\ln\left(\xi_{h}/a\right)$.

We will utilize the Green's functions formulation of the superconductivity theory \cite{AGDBook,kopninBook2001}.
For investigation of scattering by the magnetic fluctuations, we have
to operate with the matrix $4\times4$ Green's function\cite{MakiInParks1969},
\[
\hat{G}(1,2)=-\begin{pmatrix}\left\langle T_{\tau}a_{\alpha}^{\dagger}(1)a_{\beta}(2)\right\rangle  & \left\langle T_{\tau}a_{\alpha}(1)a_{\beta}(2)\right\rangle \\
\left\langle T_{\tau}a_{\alpha}^{\dagger}(1)a_{\beta}^{\dagger}(2)\right\rangle  & \left\langle T_{\tau}a_{\alpha}(1)a_{\beta}^{\dagger}(2)\right\rangle 
\end{pmatrix}.
\]
We will expand it with respect to the fluctuating exchange field.
The unperturbed Green's function can be written as
\begin{align}
\hat{G}_{0} & =-\frac{\left(i\omega_{n}\hat{\tau}_{0}+\xi_{\mathbf{p}}\hat{\tau}_{z}\right)\hat{\sigma}_{0}-\Delta\hat{\sigma}_{y}\hat{\tau}_{y}}{\omega_{n}^{2}+\xi_{\mathbf{p}}^{2}+\Delta^{2}}\label{eq:G0}
\end{align}
where $\hat{\sigma}_{a}$ and $\hat{\tau}_{b}$ are the Pauli matrices
in the spin and Nambu space, respectively. We see that the unperturbed
Green's function can be expanded as $\hat{G}=\sum_{ab}\hat{\sigma}_{a}\hat{\tau}_{b}G_{ab}$ and,
without the uniform exchange field, the only nonzero components are
$00$, $0z$, and $yy$. For the single-band BCS model, the gap equation
is
\begin{equation}
\Delta=UT\sum_{\omega_{n}}\int\frac{d^{3}\mathbf{p}}{(2\pi)^{3}}G_{yy}(\mathbf{p}),\label{eq:GapEq}
\end{equation}
where $U$ is the pairing interaction.

\section{Scattering by fluctuating exchange field\label{sec:Scattering}}

The Green's function renormalized by scattering is
\begin{align}
\hat{G}^{-1} & =\hat{G}_{0}^{-1}-\hat{\Sigma}\label{eq:renormG}
\end{align}
where $\hat{G}_{0}^{-1}\!=i\omega_{n}\hat{\sigma}_{0}\hat{\tau}_{0}\!-\xi_{\mathbf{p}}\hat{\sigma}_{0}\hat{\tau}_{z}\!+\Delta\hat{\sigma}_{y}\hat{\tau}_{y}$
and 
\begin{equation}
\hat{\Sigma}(\mathbf{p})=\int\frac{d^{3}\mathbf{q}}{(2\pi)^{3}}\left\langle \left|\tilde{h}_{\boldsymbol{q},i}\right|^{2}\right\rangle \hat{\alpha}_{i}\hat{G}(\mathbf{p}+\mathbf{q})\hat{\alpha}_{i}\label{eq:SelfEnGen}
\end{equation}
is the self-energy due to the scattering on the fluctuating exchange
field with $\hat{\boldsymbol{\alpha}}=\left(\hat{\tau}_{z}\hat{\sigma}_{x},\hat{\tau}_{0}\hat{\sigma}_{y},\hat{\tau}_{z}\hat{\sigma}_{z}\right)$\cite{MakiInParks1969}.
Using the expansion $\hat{\Sigma}(\mathbf{p})=\sum_{a,b}\Sigma_{ab}\hat{\sigma}_{a}\hat{\tau}_{b}$,
we obtain that the relevant components with $ab=00,\,0z,\,yy$ are
\begin{align*}
\Sigma_{ab}(\mathbf{p}) & =\int\frac{d^{3}\mathbf{\mathbf{p}^{\prime}}}{(2\pi)^{3}}\left\langle \left|\tilde{\boldsymbol{h}}_{\boldsymbol{p}-\mathbf{p}^{\prime}}\right|^{2}\right\rangle G_{ab}(\mathbf{p}^{\prime}).
\end{align*}
with \begin{subequations}
\begin{align}
\Sigma_{00}(\mathbf{p}) & =-\int\frac{d^{3}\mathbf{p}^{\prime}}{(2\pi)^{3}}\left\langle \left|\tilde{\boldsymbol{h}}_{\boldsymbol{p}-\boldsymbol{p}^{\prime}}\right|^{2}\right\rangle \frac{i\omega_{n}}{\omega_{n}^{2}+\xi_{\mathbf{p}^{\prime}}^{2}+\Delta^{2}},\label{eq:SefEn}\\
\Sigma_{0z}(\mathbf{p}) & =-\int\frac{d^{3}\mathbf{p}^{\prime}}{(2\pi)^{3}}\left\langle \left|\tilde{\boldsymbol{h}}_{\boldsymbol{p}-\boldsymbol{p}^{\prime}}\right|^{2}\right\rangle \frac{\xi_{\mathbf{p}^{\prime}}}{\omega_{n}^{2}+\xi_{\mathbf{p}^{\prime}}^{2}+\Delta^{2}},
\end{align}
\end{subequations}and $\Sigma_{yy}(\mathbf{p})=-\frac{\Delta}{i\omega_{n}}\Sigma_{00}(\mathbf{p})$. 

The behavior of $\hat{\Sigma}_{\mathbf{p}}$ depends on the relation
between three length scales: the magnetic correlation length $\xi_{h}$,
in-plane coherence length $\xi_{s}$, and inverse Fermi wave vector
$k_{F}^{-1}$. Consider first limiting cases qualitatively. For very
long correlations $\xi_{h}>\xi_{s}$, we have a slowly varying exchange
field. In this case, we can neglect $\mathbf{p}^{\prime}$ dependence
everywhere except $\langle |\tilde{\boldsymbol{h}}_{\boldsymbol{p}-\boldsymbol{p}^{\prime}}|^{2}\rangle $
giving 
\begin{align}
\hat{\Sigma}(\mathbf{p}) & \approx h_{0}^{2}\hat{G}_{0}(\mathbf{p}).\label{eq:SelfEnLongCorr}
\end{align}
This corresponds to the correction due to the uniform exchange field
equal to $h_{0}$ averaged over its directions. We make two observations
from this simple result, which will be essential in the further consideration:
(i) $\hat{\Sigma}(\mathbf{p})$ has the same nonzero components as
$\hat{G}_{0}(\mathbf{p})$, i.e., $00$, $yy$, and $0z$ and (ii)
the momentum dependence in $\hat{\Sigma}(\mathbf{p})$ can not be
neglected.

In the case $\xi_{h}\!<\!\xi_{s}$, we can integrate over $\xi_{\mathbf{p}^{\prime}}$
and obtain the well-known Abrikosov-Gor'kov magnetic-scattering result
\cite{AbrGorJETP61}, 
\begin{align}
\hat{\Sigma}(\mathbf{p}) & \approx\frac{1}{2\tau_{m}}\frac{-i\omega_{n}\hat{\sigma}_{0}\hat{\tau}_{0}+\Delta\hat{\sigma}_{y}\hat{\tau}_{y}}{\sqrt{\omega_{n}^{2}+\Delta^{2}}}\label{eq:SelfEnScat}
\end{align}
with the scattering rate 
\begin{align}
\frac{1}{2\tau_{m}} & =\int\frac{\pi dS_{F}^{\prime}}{(2\pi)^{3}v_{F}^{\prime}}\left\langle \left|\tilde{\boldsymbol{h}}_{\boldsymbol{p}-\boldsymbol{p}^{\prime}}\right|^{2}\right\rangle ,\label{eq:ScatRateGen}
\end{align}
which accounts for possibility that the range of $\langle |\tilde{\boldsymbol{h}}_{\boldsymbol{p}-\boldsymbol{p}^{\prime}}|^{2}\rangle $
may be much smaller than the Fermi-surface size\cite{MachidaJLTP1979}.
Note that, in contrast to the case of long correlations, Eq.~\eqref{eq:SelfEnLongCorr},
(i) the $\boldsymbol{p}$ dependence of $\hat{\Sigma}(\mathbf{p})$
in Eq.~\eqref{eq:SelfEnScat} can be neglected and (ii)$\Sigma_{0z}$
component can be omitted. These are standard approximations of the
AG theory. In the regime $\xi_{h}>k_{F}^{-1}$ the magnetic
fluctuations give small-angle scattering, see illustration in Fig.~\ref{Fig:FS-MagnFluctLength}(b).
The dependence of the scattering rate on the correlation length following
from Eq.~\eqref{eq:ScatRateGen} is sensitive to the dimensionality of
scattering. For three-dimensional scattering, the scattering rate increases
logarithmically with $\xi_{h}$ \cite{MachidaJLTP1979}. In our quasi-2D
case, we assume that scattering occurs in the whole range of $p_{z}-p_{z}^{\prime}$
but with small change of the in-plane momentum. In this case Eq.~\eqref{eq:ScatRateGen}
gives %
\begin{equation}
\frac{1}{2\tau_{m}}=\frac{2\pi}{s}\int\limits _{-\infty}^{\infty}\frac{\pi dq}{(2\pi)^{3}v_{F}}\frac{C_{h}sh_{0}^{2}\xi_{h}^{2}}{1+\xi_{h}^{2}q^{2}}=\frac{C_{h}h_{0}^{2}\xi_{h}}{4v_{F}}.\label{eq:ScatRateInterm}
\end{equation}
In general case, the product $C_{h}h_{0}^{2}\xi_{h}$ in this formula
and in several results below, can be directly computed from the correlation
function of the exchange field as 
\begin{equation}
C_{h}h_{0}^{2}\xi_{h}=\int_{0}^{\infty}dr\left\langle \tilde{\boldsymbol{h}}(r)\tilde{\boldsymbol{h}}(0)\right\rangle .\label{eq:ProdVsCorrFun}
\end{equation}
This relation allows evaluation of the scattering rate from the spin-spin correlation function, see Eq.\ \eqref{eq:hcorr-Scorr}, which can be computed for a particular magnetic model.  
We can see that in the quasi-2D case the scattering rate increases linearly
with $\xi_{h}$, much faster than in the 3D case \cite{MachidaJLTP1979}.
For completeness, we also present here the result for very short correlation
$\xi_{h}k_{F}<1$ when magnetic fluctuations scatter at all angles.
In this case we can replace $|\tilde{\boldsymbol{h}}_{\boldsymbol{p}-\boldsymbol{p}^{\prime}}|^{2}$
with $|\tilde{\boldsymbol{h}}_{0}|^{2}$and obtain the
Abrikosov-Gor'kov result for uncorrelated magnetic impurities 
\begin{equation}
\frac{1}{2\tau_{m}}=C_{h}\nu h_{0}^{2}s\xi_{h}^{2},\label{eq:ScatRateShort}
\end{equation}
where $\nu$ is the density of states. In particular, for quasi-2D
electronic spectrum $\nu=m/(2\pi\hbar^{2}s)$ where $m$ is the effective mass.

Away from the magnetic transition, the magnetic correlation length $\xi_h$
is of the order of separation between the magnetic moments $a$. For a
continuous magnetic transition inside the superconducting state, the magnetic
correlation length rapidly increases for $T\rightarrow T_{m}$ and
at some point exceeds the coherence length. At this crossover
the impact of magnetic fluctuations on superconductivity modifies
qualitatively. We now quantify the crossover between the regimes $\xi_{h}>\xi_{s}$,
Eq.~\eqref{eq:SelfEnLongCorr}, and $\xi_{h}<\xi_{s}$, Eqs.~\eqref{eq:SelfEnScat}
and \eqref{eq:ScatRateInterm}. It is important to note that in the
second (scattering) regime only two components of $\hat{\Sigma}$
are essential, $00$ and $yy$. In the first regime, however, also
the $0z$ component describing spectrum renormalization has to be included.
The latter component obviously also has to be taken into account in
the description of the crossover. First, we consider the $00$ component
(the $00$ and $yy$ components are related as $\Sigma_{yy}=-\frac{\Delta}{i\omega_{n}}\Sigma_{00}$).
As the scattering in the regime $k_{F}\xi_{h}\gg$1 is small angle,
we need to consider only a small region at the Fermi surface near
the initial momentum $\boldsymbol{p}$. Selecting the $x$ axis along
this momentum and $y$ axis in the perpendicular direction [see Fig.~\ref{Fig:FS-MagnFluctLength}(b)]
and using $\langle |\tilde{\boldsymbol{h}}_{\mathbf{\mathbf{q}}}|^{2}\rangle $
in Eq.~\eqref{eq:FlExFl-Fourier}, we transform Eq.~\eqref{eq:SefEn}
as
\begin{align}
\Sigma_{00}(\mathbf{p}) & =\!-\!\int\frac{dp_{x}^{\prime}dp_{y}^{\prime}}{(2\pi)^{2}}\frac{C_{h}h_{0}^{2}\xi_{h}^{2}}{1\!+\!\xi_{h}^{2}\left(p_{x}^{\prime}\!-\!p_{x}\right)^{2}\!+\!\xi_{h}^{2}p_{y}^{\prime}{}^{2}}\nonumber \\
\times & \frac{i\omega_{n}}{\omega_{n}^{2}\!+\!v_{F}^{2}\left(p_{x}^{\prime}\!-\!p_{F}\right)^{2}\!+\!\Delta^{2}}.\label{eq:Z00Formula}
\end{align}
Integrating with respect to $p_{y}^{\prime}$, we obtain
\begin{align}
\Sigma_{00}(\mathbf{p}) & =-\frac{C_{h}h_{0}^{2}\xi_{h}}{4\pi}\int_{-\infty}^{\infty}dp_{x}^{\prime}
\frac{1}{\sqrt{1+\xi_{h}^{2}{p_{x}^{\prime}}^{2}}}\nonumber \\
\times & \frac{i\omega_{n}}{\omega_{n}^{2}\!+\!v_{F}^{2}\left(p_{x}^{\prime}\!+\!\delta p_{x}\right)^{2}\!+\!\Delta^{2}}\nonumber \\
 & =-\frac{C_{h}h_{0}^{2}}{4\pi}\frac{i\omega_{n}}{\omega_{n}^{2}+\Delta^{2}}U\left(\delta k_{x},g_{n}\right),\label{eq:Z00Result}
\end{align}
where $\delta p_{x}=p_{x}-p_{F}$,
\begin{equation}
g_{n}=\frac{v_{F}/\xi_{h}}{\sqrt{\omega_{n}^{2}\!+\!\Delta^{2}}},\,\delta k_{x}=\!\frac{v_{F}\left(p_{x}\!-\!p_{F}\right)}{\sqrt{\omega_{n}^{2}+\Delta^{2}}}=\!\frac{\xi_{p}}{\sqrt{\omega_{n}^{2}\!+\!\Delta^{2}}},\label{eq:Def-gn}
\end{equation}
and the reduced function $U\left(k,g\right)$ is defined by the integral
\[
U\left(k,g\right)=\int\limits _{-\infty}^{\infty}du\frac{1}{\sqrt{1+u^{2}}}\frac{1}{1+\left(gu+k\right)^{2}},
\]
which can be taken analytically giving 
\begin{align}
U\left(k,g\right) & =\mathrm{Re}\left[W\left(k,g\right)\right],\label{eq:UResult}\\
W\left(k,g\right) & =\!\frac{2}{\sqrt{\left(ik\!+\!1\right)^{2}\!-\!g^{2}}}\ln\!\left(\frac{ik\!+\!1\!+\!\sqrt{\left(ik\!+\!1\right)^{2}\!-\!g^{2}}}{g}\right).\nonumber 
\end{align}
We note that the $k$ dependence of the function $U\left(k,g\right)$
corresponding to the $\xi_{p}$ dependence of the self energy is essential
only for $g\lesssim1$. The value of the function $U\left(k,g\right)$
at $k=0$ has the simple analytical form
\begin{align*}
U\left(0,g\right) & =
\begin{cases}
\frac{2}{\sqrt{1-g^{2}}}\ln\frac{1+\sqrt{1-g^{2}}}{g} & \mathrm{for}\:g<1\\
\frac{2}{\sqrt{g^{2}-1}}\arctan\sqrt{g^{2}-1} & \mathrm{for}\:g>1
\end{cases}.
\end{align*}
The asymptotic $U\left(0,g\right)\simeq \pi/g$ for $g\gg1$ corresponds
the scattering regime, Eqs.~\eqref{eq:SelfEnScat} and \eqref{eq:ScatRateInterm}.
In this limit the $k$ dependence of the function $U\left(k,g\right)$
can be neglected. On the other hand, the asymptotic for $g\ll1$ is \[U\left(k,g\right)\simeq2\left[\frac{1}{1\!+\!k^{2}}\ln\left(2\frac{\sqrt{1\!+\!k^{2}}}{g}\right)+\frac{k}{1\!+\!k^{2}}\arctan k\right].\]
It corresponds to the uniform-field result in Eq.~\eqref{eq:SelfEnLongCorr}
only for the main logarithmic term. Additional terms appear because
the correlation function $\langle \tilde{h}_{\alpha}(\boldsymbol{r})\tilde{h}_{\beta}(\boldsymbol{r}^{\prime})\rangle $
in Eq.~\eqref{eq:FlExFCorr} is not a constant at $|\boldsymbol{r}-\boldsymbol{r}^{\prime}|\!<\!\xi_{h}$
but increases logarithmically as $\ln\left(\xi_{h}/|\boldsymbol{r}-\boldsymbol{r}^{\prime}|\right)$.

As mentioned above, for the proper description of the crossover at
$\xi_{h}\sim\xi_{s}$, we also need to take into account the $0z$
component of the self energy, 
\begin{align}
\Sigma_{0z}(\mathbf{p}) & =\!-\!\int\frac{dp_{x}^{\prime}dp_{y}^{\prime}}{(2\pi)^{2}}\frac{C_{h}h_{0}^{2}\xi_{h}^{2}}{1\!+\!\xi_{h}^{2}\left(p_{x}^{\prime}\!-\!p_{x}\right)^{2}\!+\!\xi_{h}^{2}p_{y}^{\prime}{}^{2}}\nonumber \\
\times & \frac{v_{F}\left(p_{x}^{\prime}-p_{F}\right)}{\omega_{n}^{2}\!+\!v_{F}^{2}\left(p_{x}^{\prime}\!-\!p_{F}\right)^{2}\!+\!\Delta^{2}}.\label{eq:Z0zFormula}
\end{align}
Following the same route as in derivation of Eq.~\eqref{eq:Z00Result},
we present it as
\begin{align}
\Sigma_{0z}(\mathbf{p}) & =-\frac{C_{h}h_{0}^{2}}{4\pi}\frac{1}{\sqrt{\omega_{n}^{2}+\Delta^{2}}}V\left(\delta k_{x},g_{n}\right),\label{eq:Z0zResult}
\end{align}
where the parameters $g_{n}$ and $\delta k_{x}$ are defined in Eq.~\eqref{eq:Def-gn},
\begin{equation}
V\left(k,g\right)\!=\!\int\limits _{-\infty}^{\infty}\!du\frac{1}{\sqrt{1\!+\!u^{2}}}\frac{gu+k}{1\!+\!\left(gu\!+\!k\right)^{2}}=\!-\mathrm{Im}\left[W\left(k,g\right)\right],\label{eq:VResult}
\end{equation}
and the function $W\left(k,g\right)$ is defined in Eq.~\eqref{eq:UResult}.
In particular, for $g\rightarrow0$
\[
V\left(k,g\right)\simeq\frac{2}{1+k^{2}}\left[k\ln\left(\frac{2}{g}\right)-\arctan k\right].
\]
As follows from Eq.~\eqref{eq:renormG}, the renormalized Green's
function can be obtained by substitutions $i\omega_{n}\rightarrow i\tilde{\omega}_{n}=i\omega_{n}-\Sigma_{00}$,
$\Delta\rightarrow\tilde{\Delta}=\Delta-\Sigma_{yy}$, and $\xi_{\mathbf{p}}\rightarrow\tilde{\xi}_{\mathbf{p}}=\xi_{\mathbf{p}}+\Sigma_{0z}$.
The renormalized frequency, gap, and spectrum can be written as 
\begin{equation}
\tilde{\omega}_{n}  =\omega_{n}(1+\alpha_{n}),\,\tilde{\Delta}=\Delta_{0}(1-\alpha_{n}),\,\tilde{\xi}_{\mathbf{p}}=\xi_{\mathbf{p}}(1-\beta_{n})
\label{eq:tilde}
\end{equation}
with 
\begin{equation}
\alpha_{n}=\frac{C_{h}h_{0}^{2}}{4\pi}\frac{U\left(\delta k_{x},g_{n}\right)}{\omega_{n}^{2}+\Delta_{0}^{2}},\ 
\beta_{n}=\frac{C_{h}h_{0}^{2}}{4\pi}\frac{V\left(\delta k_{x},g_{n}\right)/\xi_{\mathbf{p}}}{\sqrt{\omega_{n}^{2}+\Delta_{0}^{2}}}.
\label{eq:alphabeta}
\end{equation}
With derived results for the self-energy in Eqs.~\eqref{eq:Z00Result}
and \eqref{eq:Z0zResult}, we proceed with evaluation of correction
to the gap parameter from Eq.~\eqref{eq:GapEq}. 

\section{Correction to the gap\label{sec:Gap}}

The superconducting gap is the most natural parameter characterizing the strength of superconductivity at a given temperature. 
In this section, we calculate the suppression of this parameter
by correlated magnetic fluctuations. The key observation is that a
small uniform exchange field has no influence on the gap at zero temperature
\cite{SarmaJPCS1963}. Therefore, one can expect that the suppression
caused by correlated magnetic fluctuations at low temperatures diminishes
when the magnetic correlation length exceeds the superconducting coherence
length. 

The gap equation in Eq.~\eqref{eq:GapEq} is determined by the integral
\begin{equation}
\mathcal{I}  =\int\frac{d^{3}\mathbf{p}}{(2\pi)^{3}}G_{yy}(\mathbf{p}) =\nu\int_{-\infty}^{\infty}d\xi\frac{\tilde{\Delta}}{\tilde{\omega}_{n}^{2}\!+\!\tilde{\xi}^{2}\!+\!\tilde{\Delta}^{2}},\label{eq:GapInt}
\end{equation}
where the parameters with ``$\sim$'' are defined in Eqs.\ \eqref{eq:tilde} and \eqref{eq:alphabeta}.
We evaluate the linear correction to $\mathcal{I}$ with respect to $\alpha_{n},\beta_{n}\propto h_{0}^{2}$ as 
\begin{align*}
\mathcal{\delta I} & =-\nu\int\limits _{-\infty}^{\infty}\!d\xi\frac{\Delta_{0}}{\omega_{n}^{2}\!+\!\xi^{2}\!+\!\Delta_{0}^{2}}\\
\times & \left(\alpha_{n}\!+\!\frac{2\alpha_{n}\left(\omega_{n}^{2}\!-\!\Delta_{0}^{2}\right)\!-\!2\beta_{n}\xi^{2}}{\left(\omega_{n}^{2}+\xi^{2}+\Delta_{0}^{2}\right)}\right).
\end{align*}
Making the substitution $\xi=z\sqrt{\omega_{n}^{2}+\Delta_{0}^{2}}$,
we transform this correction to the following form 
\begin{align}
\delta\mathcal{I} & =-\frac{C_{h}\nu h_{0}^{2}}{4\pi}\frac{\Delta_{0}}{\left(\omega_{n}^{2}+\Delta_{0}^{2}\right)^{3/2}}\nonumber \\
\times & \mathrm{Re}\left[\int\limits _{-\infty}^{\infty}dz\frac{W\left(z,g_{n}\right)\left(\left(z-i\right)^{2}+\frac{4\omega_{n}^{2}}{\omega_{n}^{2}+\Delta_{0}^{2}}\right)}{\left(z^{2}+1\right)^{2}}\right].\label{eq:CorrGapInt}
\end{align}
Calculation of the integral described in appendix \ref{app:CalcInt}
yields the result
\begin{align}
\delta\mathcal{I} & =-C_{h}\nu h_{0}^{2}\frac{\Delta_{0}}{\left(\omega_{n}^{2}+\Delta_{0}^{2}\right)^{5/2}}\left\{ \frac{\Delta_{0}^{2}}{4-g_{n}^{2}}\right.\nonumber \\
+ & \left.\frac{\omega_{n}^{2}\left(4\!-\!g_{n}^{2}\right)\!-\!2\Delta_{0}^{2}}{\left(4-g_{n}^{2}\right)^{3/2}}\ln\left(\frac{2\!+\!\sqrt{4\!-\!g_{n}^{2}}}{g_{n}}\right)\right\} .\label{eq:CorrGapIntResult}
\end{align}
The corrected equation for the gap $\Delta=UT\sum_{\omega_{n}}(\mathcal{I}_{0}+\delta\mathcal{I})$
with $\mathcal{I}_{0}=\pi/\sqrt{\omega_{n}^{2}+\Delta_{0}^{2}}$ gives
the gap correction caused by the nonuniform exchange field
\begin{align*}
\tilde{\Delta} & =-C_{h}h_{0}^{2}T\!\sum_{\omega_{n}}\!\frac{1}{\left(\omega_{n}^{2}\!+\!\Delta_{0}^{2}\right)^{5/2}}.\\
\times & \mathrm{Re}\!\left[\frac{\Delta_{0}^{2}}{4\!-\!g_{n}^{2}}\!+\!\frac{4\omega_{n}^{2}\!-\!2\Delta_{0}^{2}\!-\!\omega_{n}^{2}g_{n}^{2}}{\left(4-g_{n}^{2}\right)^{3/2}}\ln\!\left(\!\frac{2\!+\!\sqrt{4\!-\!g_{n}^{2}}}{g_{n}}\!\right)\!\right]\\
\times & \left[\pi T\!\sum_{\omega_{n}}\frac{\Delta_{0}}{\left(\omega_{n}^{2}\!+\!\Delta_{0}^{2}\right)^{3/2}}\right]^{-1}.
\end{align*}
Substituting the definition of $g_{n}$ in Eq.~\eqref{eq:Def-gn},
we finally obtain
\begin{align}
\tilde{\Delta}&	=\!-\left[\pi T\sum_{\omega_{n}}\frac{\Delta_{0}}{\varOmega_{n}^{3}}\right]^{-1}\!
C_{h}h_{0}^{2}T\sum_{\omega_{n}}\frac{1}{\varOmega_{n}^{3}\left(4\varOmega_{n}^{2}-\varepsilon_{h}^{2}\right)}\label{eq:GapCorr}\\
\times&	\!\left[\!\Delta_{0}^{2}\!+\!\frac{2\left(2\omega_{n}^{2}\!-\!\Delta_{0}^{2}\right)\varOmega_{n}^{2}\!-\!\omega_{n}^{2}\varepsilon_{h}^{2}}{\varOmega_{n}\sqrt{4\varOmega_{n}^{2}-\varepsilon_{h}^{2}}}\ln\!\left(\!\frac{2\varOmega_{n}\!+\!\sqrt{4\varOmega_{n}^{2}\!-\!\varepsilon_{h}^{2}}}{\varepsilon_{h}}\right)\!\right]\nonumber
\end{align}
with $\varOmega_{n}=\sqrt{\omega_{n}^{2}\!+\!\Delta_{0}^{2}}$ and 
the magnetic-scattering energy scale $\varepsilon_{h}\!=\!v_{F}/\xi_{h}$. Introducing
the reduced variables 
\begin{align*}
\tilde{T}\!=&2\pi T/\Delta_{0}(T),\  
\tilde{\omega}_n\!=\!\tilde{T}(n\!+\!1/2),\\  \alpha_{h}\!=&\varepsilon_{h}(T)/2\Delta_{0}(T)\!=\!\xi_{s}(T)/\xi_{h}(T)
\end{align*}
with $\xi_{s}(T)\!=\!v_{F}/2\Delta_{0}(T)$, and using the estimate $C_{h}\!\approx \! 2\pi/\ln\left(\xi_{h}/a\right)$,
we rewrite this result in the form convenient for numerical evaluation
\begin{widetext}
\begin{subequations}
\begin{align}
\tilde{\Delta}(T)= & -\frac{h_{0}^{2}}{2\Delta_{0}(T)\ln\left(\xi_{h}(T)/a\right)}\mathcal{V}_{\Delta}\left(\frac{2\pi T}{\Delta_{0}(T)},\frac{\xi_{s}(T)}{\xi_{h}(T)}\right),\label{eq:GapCorrRed}\\
\mathcal{V}_{\Delta}\left(\tilde{T},\alpha_{h}\right) & =\left[\mathcal{D}(\tilde{T})\right]^{-1}\tilde{T}\sum_{n=0}^{\infty}
R\left[\tilde{\omega}_n,\alpha_{h}\right],\label{eq:VD-T-a}\\
\mathcal{D}(\tilde{T}) & =\tilde{T}\sum_{n=0}^{\infty}\left(\tilde{\omega}_n^{2}\!+1\right)^{-3/2},\label{eq:DenomT}\\
R\left(z,\alpha_{h}\right) & =\frac{1}{\left(z^{2}\!+\!1\right)^{3/2}\left(z^{2}\!+\!1\!-\!\alpha_{h}^{2}\right)}\left[1\!+\!\left(2z^{2}\!-\!1\!-\!\frac{2z^{2}\alpha_{h}^{2}}{z^{2}\!+\!1}\right)\mathit{L}\left(z,\alpha_{h}\right)\right].\label{eq:Rza}\\
\mathit{L}\left(z,\alpha_{h}\right) & =\begin{cases}
\frac{\sqrt{z^{2}+1}}{\sqrt{z^{2}+1-\alpha_{h}^{2}}}\ln\left(\frac{\sqrt{z^{2}\!+\!1}\!+\!\sqrt{z^{2}\!+\!1\!-\!\alpha_{h}^{2}}}{\alpha_{h}}\right), & z^{2}\!>\!\alpha_{h}^{2}\!-\!1\\
\frac{\sqrt{z^{2}+1}}{\sqrt{\alpha_{h}^{2}-z^{2}-1}}\arctan\frac{\sqrt{\alpha_{h}^{2}\!-\!z^{2}\!-\!1}}{\sqrt{z^{2}\!+\!1}}, & z^{2}\!<\!\alpha_{h}^{2}\!-\!1
\end{cases}.\label{eq:LogFun}
\end{align}
\end{subequations}
\end{widetext}
Note that the function $L\left(z,\alpha_{h}\right)$
does not have singularity at $z=\sqrt{\alpha_{h}^{2}-1}$ for $\alpha_{h}>1$,
contrary to what its shape may suggest. We see that the gap correction
has the amplitude $h_{0}^{2}/\Delta_{0}$ and mostly depends on two
dimensionless parameters: reduced temperature $T/\Delta_{0}(T)$ and
the ratio $\alpha_h\!=\!\xi_{s}(T)/\xi_{h}(T)$.
It also weakly depends on the ratio $\xi_{h}/a$, which determines
the logarithmic factor in the denominator of Eq.~\eqref{eq:GapCorrRed}.

Let us discuss asymptotic behavior of the reduced function $\mathcal{V}_{\Delta}(\tilde{T},\alpha_{h})$
and the gap correction it gives. In the range $\alpha_{h}\gg1$ corresponding
to the scattering regime, the function $R\left(z,\alpha_{h}\right)$
in Eq.~\eqref{eq:Rza} behaves as $R\left(z,\alpha_{h}\right)\simeq\frac{\pi z^{2}}{\alpha_{h}\left(z^{2}+1\right)^{2}}$.
This gives the following asymptotics of the function $\mathcal{V}_{\Delta}(\tilde{T},\alpha_{h})$
\begin{subequations}
\begin{align}
\mathcal{V}_{\Delta}\left(\tilde{T},\alpha_{h}\right) & \simeq\frac{\pi}{\alpha_{h}}V_{\Delta}(\tilde{T}),\label{eq:VD-Asymp}\\
V_{\Delta}(\tilde{T})= & \left[\mathcal{D}(\tilde{T})\right]^{-1}\!\tilde{T}\sum_{n=0}^{\infty}\frac{\tilde{\omega}_n^2}{\left(\tilde{\omega}_n^2+1\right)^{2}},\label{eq:VDScatt}
\end{align}
\end{subequations}
where the limiting behaviors of the function $V_{\Delta}(\tilde{T})$
are $V_{\Delta}(0)\!=\!\pi/4$ and $V_{\Delta}(\tilde{T})\!\simeq\!{\pi^{2}\tilde{T}}/[{14\zeta(3)}]$ for
$\tilde{T}\!\gg\!1$ with $\zeta(3)\approx 1.202$. Correspondingly, the correction to the gap in this
regime simplifies to
\begin{equation}
\tilde{\Delta}(T)\simeq-\frac{\pi\xi_{h}h_{0}^{2}}{v_{F}\ln\left(\xi_{h}/a\right)}V_{\Delta}(\tilde{T})=-\frac{1}{\tau_{m}}V_{\Delta}(\tilde{T}).\label{eq:GapCorrMagnScat}
\end{equation}
This is a well-known result for the gap correction caused by the magnetic
scattering\cite{AbrGorJETP61,SkalskiPhysRev.136.A1500}.

\begin{figure}
\includegraphics[width=3.1in]{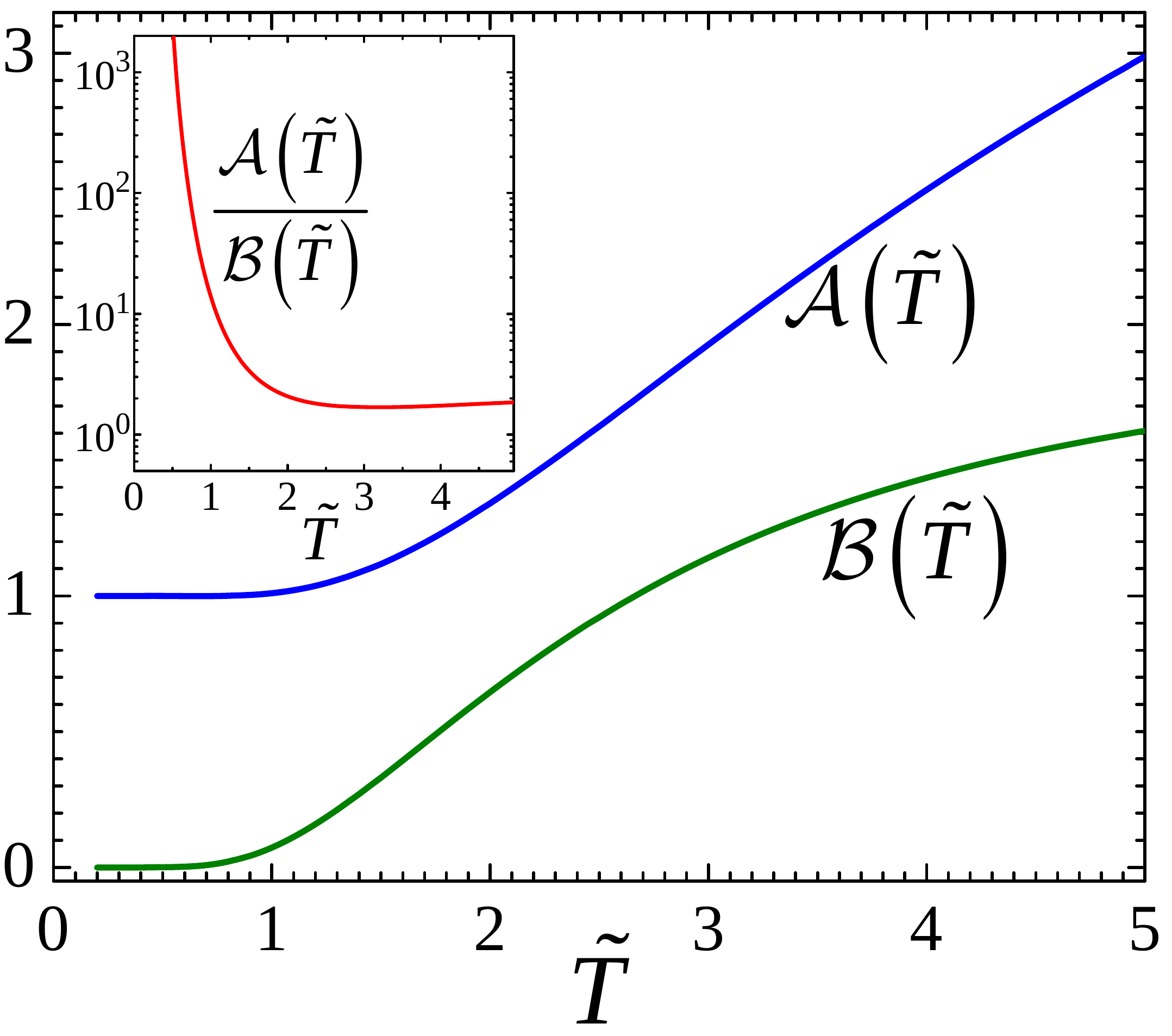}
\caption{Temperature dependences of the coefficients $\mathcal{A}(\tilde{T})$
and $\mathcal{B}(\tilde{T})$ which determine the small-$\alpha_{h}$
asymptotics of the function $\mathcal{V}_{\Delta}(\tilde{T},\alpha_{h})$
in Eq.~\eqref{eq:VDsmall-ah}. The inset shows the temperature dependence
of the ratio $\mathcal{A}(\tilde{T})/\mathcal{B}(\tilde{T})$.\label{fig:Delta-Small-ah}}
\end{figure}

In the opposite regime $\alpha_{h}\rightarrow0$ corresponding to
the vicinity of the magnetic transition, the function $R\left(z,\alpha_{h}\right)$
has a logarithmic dependence on $\alpha_{h}$ 
\begin{align*}
R\left(z,\alpha_{h}\right) & =R_{0}(z)+R_{1}(z)\ln\left(\frac{1}{\alpha_{h}}\right),\\
R_{0}(z)= & \frac{1\!+\!\left(2z^{2}\!-\!1\right)\ln\left(2\sqrt{z^{2}\!+\!1}\right)}{\left(z^{2}+1\right)^{5/2}},\\
R_{1}(z)= & \frac{2z^{2}-1}{\left(z^{2}+1\right)^{5/2}}
\end{align*}
meaning that $\mathcal{V}_{\Delta}(\tilde{T},\alpha_{h})$
also logarithmically diverges for $\alpha_{h}\rightarrow0$,
\begin{subequations}
\begin{align}
\mathcal{V}_{\Delta}\left(\tilde{T},\alpha_{h}\right) & =\mathcal{A}(\tilde{T})+\mathcal{B}(\tilde{T})\ln\left(\frac{1}{\alpha_{h}}\right),
\label{eq:VDsmall-ah}\\
\mathcal{A}(\tilde{T}) & =\left[\mathcal{D}(\tilde{T})\right]^{-1}\!\tilde{T}\sum_{n=0}^{\infty}\!
R_{0}(\tilde{\omega}_n),
\label{eq:AVDsmall-ah} \\
\mathcal{B}(\tilde{T}) & =\left[\mathcal{D}(\tilde{T})\right]^{-1}\tilde{T}\sum_{n=0}^{\infty}
R_{1}(\tilde{\omega}_n)
\label{eq:BVDsmall-ah}
\end{align}
\end{subequations}
with $\mathcal{A}(0)=1$, $\mathcal{B}(0)=0$.
The plots of the coefficients $\mathcal{A}(\tilde{T})$
and $\mathcal{B}(\tilde{T})$ are shown in Fig.~\ref{fig:Delta-Small-ah}.
The corresponding correction to the gap can be presented as 
\begin{align}
\tilde{\Delta}(\tilde{T}) & =\!-\frac{h_{0}^{2}\mathcal{B}(\tilde{T})}{2\Delta_{0}}\left[1\!-\frac{\ln\left(\frac{\xi_{s}}{a}\right)\!-\!\mathcal{A}(\tilde{T})\!/\!\mathcal{B}(\tilde{T})}{\ln\left(\xi_{h}/a\right)}\right].\label{eq:DeltaLarge-xih}
\end{align}
Therefore, the absolute value of correction $|\tilde{\Delta}|$
decreases when $T$ approaches $T_{m}$ if the ratio $\mathcal{A}(\tilde{T}_{m})/\mathcal{B}(\tilde{T}_{m})$
exceeds $\ln\left(\xi_{s}/a\right)$, which always occurs at sufficiently
low temperatures, see inset in Fig.~\ref{fig:Delta-Small-ah}. In
this case the overall dependence of the correction on $\xi_{h}$ is
nonmonotonic and maximum suppression of the gap occurs at $\xi_{h}\sim\xi_{s}$.
The limiting value at $T=T_{m}$ , $\tilde{\Delta}(\tilde{T}_{m})=-h_{0}^{2}\mathcal{B}(\tilde{T}_{m})/2\Delta_{0}$,
corresponds to the correction from a uniform exchange field equal
to $h_{0}$. It vanishes for $T_{m}\rightarrow0$ as $\exp\left(-\Delta_{0}/T_{m}\right)$.

At temperatures much smaller than $T_{c}$, the summation over the
Matsubara frequencies in Eq.~\eqref{eq:VD-T-a} can be transformed
into integration leading to
\begin{align}
\tilde{\Delta}(0) & =-\frac{h_{0}^{2}}{2\Delta_{0}\ln\left(\xi_{h}/a\right)}
\mathcal{V}_{\Delta}\!\left(\frac{\xi_{s}}{\xi_{h}}\right),
\label{eq:dGapT0}
\end{align}
where the reduced function $\mathcal{V}_{\Delta}\left(\alpha_{h}\right)\equiv\mathcal{V}_{\Delta}\left(0,\alpha_{h}\right)$
is defined by the integral
\begin{align*}
\mathcal{V}_{\Delta}\left(\alpha_{h}\right) & =\int_{0}^{\infty}R\left(z,\alpha_{h}\right)dz.
\end{align*}
This is a monotonically-decreasing function with the asymptotics
\begin{equation}
\mathcal{V}_{\Delta}\left(\alpha_{h}\right)\!\simeq\!\begin{cases}
1\!+\frac{1}{9}(6\ln\alpha_{h}\!+\!1)\alpha_{h}^{2}, & \mathrm{for}\,\alpha_{h}\!\ll\!1\\
\frac{\pi^{2}}{4\alpha_{h}}-\frac{2\ln\alpha_{h}+1}{\alpha_{h}^{2}}, & \mathrm{for}\,\alpha_{h}\!\gg\!1
\end{cases}.\label{eq:VD-asymp}
\end{equation}
It also has the exact value $\mathcal{V}_{\Delta}(1)=\frac{\pi^{2}}{8}-\frac{1}{2}$.
The large-$\alpha_{h}$ asymptotics corresponds to the magnetic-scattering
regime\cite{AbrGorJETP61,SkalskiPhysRev.136.A1500,MakiInParks1969}.
Substituting the first leading term into Eq.~\eqref{eq:dGapT0} yields
the known result for the gap correction at zero temperature $\tilde{\Delta}(0)\approx-\pi/4\tau_{m}$,
where $\tau_{m}$ is given by Eq.~\eqref{eq:ScatRateInterm}.

\begin{figure}
\includegraphics[width=3.4in]{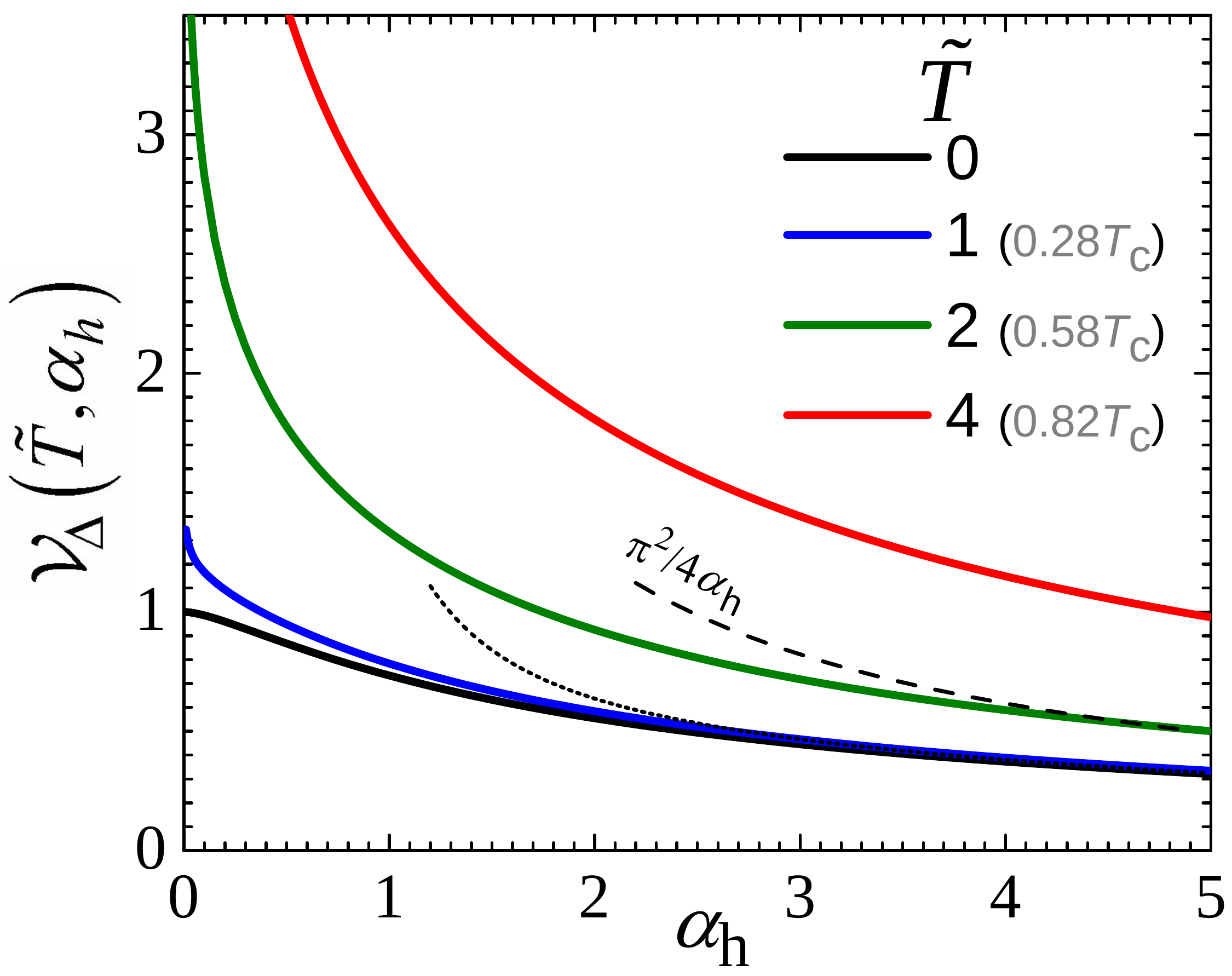}\caption{Plots of the function $\mathcal{V}_{\Delta}(\tilde{T},\alpha_{h})$
in Eq.~\eqref{eq:VD-T-a} determining the gap correction caused by
nonuniform exchange field on the parameter $\alpha_{h}=\xi_{s}/\xi_{h}$
for several values of the reduced temperature $\tilde{T}=2\pi T/\Delta_{0}$.
The corresponding relative temperatures for BCS superconductors are
shown in parenthesis. For zero temperature, we also show the scattering-regime
asymptotics (dashed line) and more accurate asymptotics presented
in Eq.~\eqref{eq:VD-asymp} (dotted line).}
\label{Fig:VD-a-T}
\end{figure}

Plots of the numerically evaluated function $\mathcal{V}_{\Delta}(\tilde{T},\alpha_{h})$
are shown in Fig,~\ref{Fig:VD-a-T} for several values of the reduced
temperature $\tilde{T}$. The function monotonically decreases with
$\alpha_{h}$ and increases with temperature.
At zero temperature this function approaches a finite value for $\alpha_{h}\rightarrow0$
while at finite temperatures it logarithmically diverges, as discussed above.
For zero temperature, we also show the scattering-regime dependence
by dashed line and more accurate asymptotic presented in Eq.~\eqref{eq:VD-asymp}
by dotted line. We can see that the scattering approximation noticeably
overestimates the gap correction for rather large values of $\alpha_{h}$.
The finite value of the function for $\alpha_{h}\rightarrow0$ at
zero temperature is in an apparent contradiction with the known result
that a uniform exchange field does not change the gap at zero temperature
\cite{SarmaJPCS1963}. This finite value is the consequence of small-distance
behavior of the exchange-field correlation function for the two-dimensional
case: it does not approach a constant for $r\ll\xi_{h}$ but keeps
growing logarithmically, see Eqs.~\eqref{eq:FlExFCorr} and \eqref{eq:fhr}.
We note, however, that despite this small-$\alpha_{h}$ saturation
of the function $\mathcal{V}_{\Delta}\left(0,\alpha_{h}\right)$,
the gap correction in Eq.~\eqref{eq:GapCorrRed} does have a nonmonotonic
dependence on $\xi_{h}$ and vanishes in the limit $\xi_{h}\rightarrow\infty$
at low temperatures because of the logarithmic factor in the denominator.

\section{Correction to the electromagnetic kernel and London penetration depth\label{sec:Kernel}}

In this section, we investigate the correction to the superconducting
current response caused by the exchange interaction with correlated
magnetic fluctuations. As in the case of the gap parameter, there
are two different regimes depending on the relation between the magnetic
correlation length $\xi_{h}$ and superconducting coherence length
$\xi_{s}$. Our goal is to quantitatively describe the crossover between
these two regimes. The case $\xi_{h}<\xi_{s}$ corresponds to the
well-studied magnetic-scattering regime. Influence of magnetic scattering
on the electromagnetic kernel, which determines the London penetration
depth, was investigated by Skalski \emph{et al}.\cite{SkalskiPhysRev.136.A1500},
see also Ref.~\cite{MakiInParks1969}. Recently, a very detailed
investigation of this problem has been performed within the quasiclassical
approach \cite{KoganLonPenDepthPhysRevB13}. Most studies, however,
have been done for isotropic magnetic scattering. The case of correlated
magnetic fluctuation in the regime $k_{F}\xi_{h}\gg1$ requires a proper
accounting for the vertex correction to the kernel which is equivalent
to accounting for the reverse scattering events in quasiclassical approach
\footnote{The vertex correction for arbitrary magnetic scattering has been considered
in Ref.~\cite{SkalskiPhysRev.136.A1500}. The recipe to account
for the vertex correction in the kernel given after Eq.~(6.9), however,
contains a mistake: the sign in front of $\Gamma^{t}$ is incorrect.
}.

The superconducting current response
\begin{equation}
j_{\alpha}(\boldsymbol{q},\omega)=-Q_{\alpha\beta}(\boldsymbol{q},\omega)A_{\beta}(\boldsymbol{q},\omega)\label{eq:CurrentResp}
\end{equation}
is determined by the electromagnetic kernel $Q_{\alpha\beta}(\boldsymbol{q},\omega)$.
In strongly type-II superconductors, the screening of magnetic field
is determined by the local static kernel $Q_{\alpha\beta}\!\equiv\!Q_{\alpha\beta}(\boldsymbol{0},0)$.
The superfluid density $n_{s}$ introduced in the phenomenological
London theory is related to $Q_{\alpha\beta}$ as $Q_{\alpha\beta}\!=\!e^{2}n_{s}/cm_{\alpha\beta}$,
where $m_{\alpha\beta}$ is the effective mass tensor. The London
penetration depth components $\lambda_{\alpha}$ are related to the
static uniform kernel as $Q_{\alpha\alpha}\!=\!c/(4\pi\lambda_{\alpha}^{2})$.
In nonmagnetic superconductors the screening length $\tilde{\lambda}_{\alpha}$
is identical to this 'bare' length $\lambda_{\alpha}$ defined via
the electromagnetic kernel. In magnetic superconductors, however,
the screening length $\tilde{\lambda}_{\alpha}$ is reduced by the
magnetic response of local moments as $\tilde{\lambda}_{\alpha}\!=\!\lambda_{\alpha}/\sqrt{\mu}$,
where $\mu$ is the magnetic permeability in the magnetic-field direction
\cite{MaekawaJMMM79,BulaevskiiAdvPhys85}. 
Note that the exchange and magnetic response have opposite influences on the screening length: the former enlarges and the latter reduces it.  
In the following, we concentrate on the calculation of the bare London penetration depth.

In the Green's function formalism, the kernel can be evaluated as\cite{MakiInParks1969}
\begin{align}
Q_{\alpha\beta}(\boldsymbol{q},\omega_{\nu}) & =\frac{e^{2}n}{cm_{\alpha\beta}}+\frac{e^{2}}{2c}T\sum_{\omega_{n}}\!\int\!\frac{d^{3}\boldsymbol{p}}{(2\pi)^{3}}v_{\alpha}v_{\beta}\nonumber \\
\times & \mathrm{Tr}\left[\hat{G}(\boldsymbol{p},\omega_{n})\hat{G}(\boldsymbol{p}\!-\!\boldsymbol{q},\omega_{n}\!-\!\omega_{\nu})\right],\label{eq:KernelDef}
\end{align}
where $n$ is total density and $v_\alpha=\partial \xi_p/\partial p_\alpha$ are the velocity components \footnote{Note that $n/m_{\alpha\beta}=2\nu\left\langle v_{\alpha}v_{\beta}\right\rangle$ where $\nu$ is the density of states per spin}.
In particular, for clean case
\begin{align}
Q_{\alpha\beta}^{(0)} & =\frac{2\pi e^{2}}{c}\nu\left\langle v_{\alpha}v_{\beta}\right\rangle T\sum_{\omega_{n}}\frac{\Delta_{0}^{2}}{\left(\omega_{n}^{2}+\Delta_{0}^{2}\right)^{3/2}}\label{eq:CleanQ}
\end{align}
giving $Q_{\alpha\beta}^{(0)}=2\frac{e^{2}}{c}\nu\left\langle v_{\alpha}v_{\beta}\right\rangle $
at zero temperature.

We first consider the scattering regime, $\xi_{h}\ll\xi_{s}$, within
the quasiclassical approximation. The generalization of the isotropic-scattering
calculations in Ref.\ \cite{KoganLonPenDepthPhysRevB13} for arbitrary
scattering described in Appendix \ref{app:Magn-Scatt-Lond} gives
the following result for the correction to $\lambda_{\alpha}^{-2}$
due to the magnetic scattering
\begin{align}
\lambda_{1\alpha}^{-2}(T)= & \!-\!\lambda_{0\alpha}^{-2}(T)\left[\frac{1}{\tau_{m}\Delta_{0}(T)}V_{\lambda,m}\left(\frac{2\pi T}{\Delta_{0}(T)}\right)\right.\nonumber \\
+ & \left.\frac{1}{\tau_{m}^{\mathrm{tr}}\Delta_{0}(T)}V_{\lambda,m}^{\mathrm{tr}}\left(\frac{2\pi T}{\Delta_{0}(T)}\right)\right],\label{eq:LambCorrScatt}
\end{align}
with
\begin{align}
V_{\lambda,m}\!(\tilde{T}) & \!=\!\frac{1}{\mathcal{D}(\tilde{T})}
\tilde{T}\!\sum_{n=0}^{\infty}\left[\!\frac{2\tilde{\omega}_n^2\!-\!1}
{\left(1\!+\!\tilde{\omega}_n^2\right)^{5/2}}V_{\Delta}\!(\tilde{T}) 
\!+\!\frac{3\tilde{\omega}_n^2-1}{\left(1\!+\!\tilde{\omega}_n^2\right)^{3}}\right],
 \label{eq:Vlm}\\
V_{\lambda,m}^{\mathrm{tr}}\!(\tilde{T}) & \!=\frac{1}{2\mathcal{D}(\tilde{T})}\tilde{T}\sum_{n=0}^{\infty}\frac{1-\tilde{\omega}_n^2}{\left(1+\tilde{\omega}_n^2\right)^{3}},
\end{align}
where the functions $\mathcal{D}(\tilde{T})$ and $V_{\Delta}(\tilde{T})$
are defined in Eqs.~\eqref{eq:DenomT} and \eqref{eq:VDScatt}, respectively.
Here $\tau_{m}$ is the magnetic-scattering lifetime, Eqs.~\eqref{eq:ScatRateGen}
and \eqref{eq:ScatRateInterm}, and $\tau_{m}^{\mathrm{tr}}$ is the
corresponding transport time,
\begin{align}
\frac{1}{2\tau_{m}^{\mathrm{tr}}} & =\int\frac{\pi dS_{F}^{\prime}}{(2\pi)^{3}v_{F}^{\prime}}\left(1-\frac{\boldsymbol{v}\cdot\boldsymbol{v}^{\prime}}{\left\langle v^{2}\right\rangle }\right)\left\langle \left|\tilde{\boldsymbol{h}}_{\boldsymbol{p}-\boldsymbol{p}^{\prime}}\right|^{2}\right\rangle .\label{eq:TrScatRate}
\end{align}
The two terms in Eq.~\eqref{eq:LambCorrScatt} can be referred to
as the pair-breaking and transport contributions. In the case we consider,
the transport scattering rate is much smaller than the total rate,
$1/\tau_{m}^{\mathrm{tr}}\!\sim\!1/\left(\xi_{h}k_{F}\tau_{m}\right)\!\ll\!1/\tau_{m}$,
and it does not increase when the temperature approaches the magnetic transition.
We point, however, that the contribution from the total scattering
rate vanishes at low temperatures, $V_{\lambda,m}(0)\!=\!0$,
while the transport contribution remains finite $V_{\lambda,m}^{\mathrm{tr}}(0)\!=\!\pi/16$.
Nevertheless, as our main goal is to understand suppression of the
superconducting parameters near the magnetic transition, in the following
consideration we mostly focus on the behavior of the pair-breaking term proportional
to the total scattering rate.

The above results are only valid until $\xi_{h}<\xi_{s}$. We proceed
with the consideration of the crossover to the opposite regime, which
can not be treated within the quasiclassical approach. The total correction
to the electromagnetic kernel is 
	\begin{figure*}
		\includegraphics[width=5.5in]{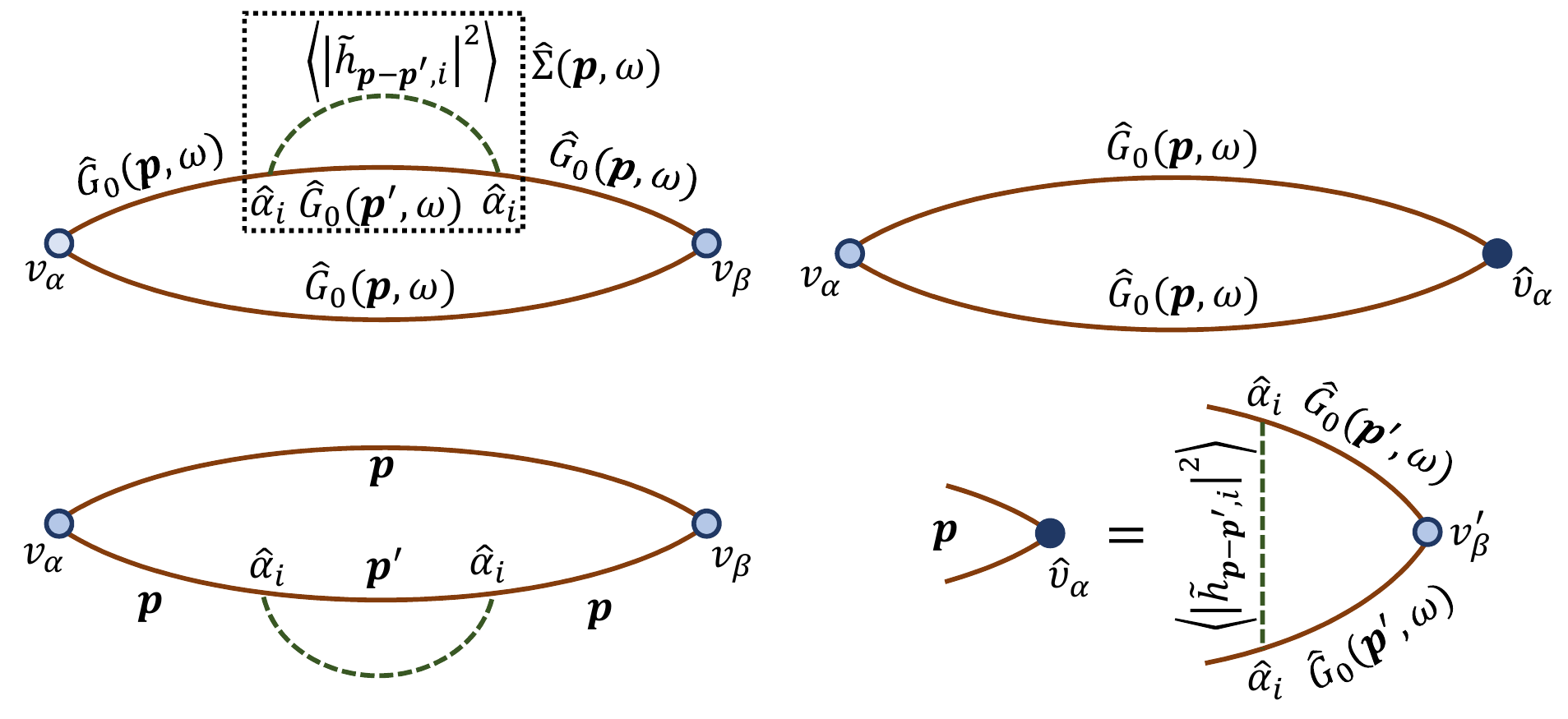}\caption{The diagrams for the lowest-order corrections to the electromagnetic kernel caused by the nonuniform exchange field in Eqs.\ \eqref{eq:KernCorrFull} and \eqref{eq:Cab}. The left-column diagrams represent the self-energy correction and the upper diagram in the right column gives the vertex correction. The lower diagram in the right columns illustrates the equation for the vertex, Eq.\  \eqref{eq:Vertex}.}
		\label{Fig:QDiagrams}
	\end{figure*}
\begin{widetext}
\begin{align}
\delta Q_{\alpha\beta} & =\frac{e^{2}}{2c}T\sum_{\omega_{n}}\mathcal{C}_{\alpha\beta}(\omega_{n}),\label{eq:KernCorrFull}\\
\mathcal{C}_{\alpha\beta}(\omega)=\! & \int\!\frac{d^{3}\boldsymbol{p}}{(2\pi)^{3}}v_{\alpha}\left\{ 2\mathrm{Tr}\!\left[\hat{G}(\boldsymbol{p},\omega)v_{\beta}\hat{G}_0(\boldsymbol{p},\omega)\hat{\Sigma}(\boldsymbol{p},\omega)\hat{G}_0(\boldsymbol{p},\omega)\right]\!+\!\mathrm{Tr}\!\left[\hat{G}_0(\boldsymbol{p},\omega)\hat{\upsilon}_{\beta}\hat{G}_0(\boldsymbol{p},\omega)\right]\right\} ,\label{eq:Cab}
\end{align}
\end{widetext}
where the first term in $\mathcal{C}_{\alpha\beta}(\omega)$ is the
self-energy correction with $\hat{\Sigma}(\boldsymbol{p},\omega)$
given by Eq.~\eqref{eq:SelfEnGen} and the second term is the vertex
correction with
\begin{align}
\hat{\upsilon}_{\beta} & =\sum_{i}\int\frac{d^{3}\boldsymbol{p}^{\prime}}{(2\pi)^{3}}\hat{\alpha}_{i}\hat{G}_{0}(\boldsymbol{p}^{\prime})v_{\beta}^{\prime}\hat{G}_{0}(\boldsymbol{p}^{\prime})\hat{\alpha}_{i}\left\langle \left|\tilde{h}_{\boldsymbol{p}-\boldsymbol{p}^{\prime},i}\right|^{2}\right\rangle .\label{eq:Vertex}
\end{align}
Figure \ref{Fig:QDiagrams} shows the diagrammatic presentation of these equations.
We split the vertex correction into two contributions 
\begin{align*}
\hat{\upsilon}_{\beta} & =v_{\beta}\hat{\Gamma}_{\boldsymbol{p}}+\delta\hat{\upsilon}_{\beta},\\
\hat{\Gamma}_{\boldsymbol{p}} & =\sum_{i}\int\frac{d^{3}\boldsymbol{p}^{\prime}}{(2\pi)^{3}}\hat{\alpha}_{i}\hat{G}_{0}(\boldsymbol{p}^{\prime})\hat{G}_{0}(\boldsymbol{p}^{\prime})\hat{\alpha}_{i}\left\langle \left|\tilde{h}_{\boldsymbol{p}-\boldsymbol{p}^{\prime},i}\right|^{2}\right\rangle ,\\
\delta\hat{\upsilon}_{\beta} & =\sum_{i}\!\int\!\frac{d^{3}\boldsymbol{p}^{\prime}}{(2\pi)^{3}}\hat{\alpha}_{i}\hat{G}_{0}(\boldsymbol{p}^{\prime})\left(v_{\beta}^{\prime}\!-\!v_{\beta}\right)\hat{G}_{0}(\boldsymbol{p}^{\prime})\hat{\alpha}_{i}\\
\times&\left\langle \left|\tilde{h}_{\boldsymbol{p}-\boldsymbol{p}^{\prime},i}\right|^{2}\right\rangle .
\end{align*}
The second contribution $\delta\hat{\upsilon}_{\beta} $ is proportional to the transport scattering rate and in our situation is typically smaller than the first one.
We therefore focus on the calculation of the first contribution. 

\begin{figure}
	\includegraphics[width=3.4in]{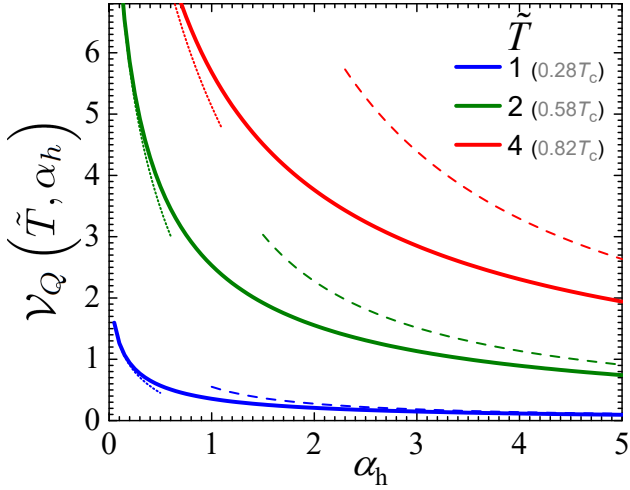}\caption{Plots of the function $\mathcal{V}_{Q}(\tilde{T},\alpha_{h})$
		in Eq.~\eqref{eq:VQ} which determines the correction to the electromagnetic
		kernel and London penetration depth in Eq.~\eqref{eq:LambCorrCorrFl}.
		The dashed lines show large-$\alpha_{h}$ asymptotics, $\mathcal{V}_{Q}(\tilde{T},\alpha_{h})\propto1/\alpha_{h}$,
		corresponding to the scattering regime. The dotted lines show small-$\alpha_{h}$
		asymptotics, $\mathcal{V}_{Q}(\tilde{T},\alpha_{h})\propto\ln\left(1/\alpha_{h}\right)$.}
	\label{Fig:VQ}
\end{figure}

Using the relations\begin{subequations}
\begin{align}
\hat{G}_{0}\hat{G}_{0} & =i\frac{\partial\hat{G}_{0}}{\partial\omega},\label{eq:GFreqDeriv}\\
\hat{\Gamma}= & i\frac{\partial\hat{\Sigma}_{\mathbf{p}}}{\partial\omega},\label{eq:WardIdent}
\end{align}
\end{subequations}where the second relation is usually called Ward
identity, we can present $\mathcal{C}_{\alpha\beta}(\omega)$ as
\begin{align}
\mathcal{C}_{\alpha\beta}(\omega) & \!=\!\mathcal{C}_{\alpha\beta}^{\mathrm{m}}(\omega)+\mathcal{C}_{\alpha\beta}^{\mathrm{tr}}(\omega),\nonumber \\
\mathcal{C}_{\alpha\beta}^{\mathrm{m}}(\omega) & \!=\!i\frac{\partial}{\partial\omega}\!
\int\!\frac{d^{3}\boldsymbol{p}}{(2\pi)^{3}}v_{\alpha}v_{\beta}\mathrm{Tr}\left[\hat{G}_0(\boldsymbol{p},\omega)\hat{\Sigma}(\boldsymbol{p},\omega)\hat{G}_0(\boldsymbol{p},\omega)\right],\label{eq:CmDef}\\
\mathcal{C}_{\alpha\beta}^{\mathrm{tr}}(\omega) & \!=\!\int\frac{d^{3}\boldsymbol{p}}{(2\pi)^{3}}v_{\alpha}\mathrm{Tr}\left[\hat{G}_0(\boldsymbol{p},\omega)\delta\hat{\upsilon}_{\beta}\hat{G}_0(\boldsymbol{p},\omega)\right].\nonumber 
\end{align}
The term $\mathcal{C}_{\alpha\beta}^{\mathrm{tr}}(\omega)$ corresponds
to contribution in Eq.~\eqref{eq:LambCorrScatt} proportional to
the transport magnetic scattering rate $1/\tau_{m}^{\mathrm{tr}}$.
As discussed above, in our case this term is typically small and does
not increase when the temperature approaches the magnetic transition.
That is why we will neglect this term in the following consideration.
As the term $\mathcal{C}_{\alpha\beta}^{\mathrm{m}}(\omega)$ is proportional
to a full derivative with respect to $\omega$, it vanishes at zero
temperature. To evaluate this term, we explicitly compute the trace
inside the integral as
\begin{align*}
&\mathrm{Tr}\left[\hat{G}_0(\boldsymbol{p},\omega)\hat{\Sigma}(\boldsymbol{p},\omega)\hat{G}_0(\boldsymbol{p},\omega)\right]  \!=\!4\left[\left(G_{00}^{2}\!+\!G_{0z}^{2}\!+\!G_{yy}^{2}\right)\Sigma_{00}\right.\\
&+\!\left.2G_{00}G_{0z}\Sigma_{0z}\!+\!2G_{00}G_{yy}\Sigma_{yy}\right],
\end{align*}
where $\Sigma_{00}$ and $\Sigma_{0z}$ are given by Eqs.~\eqref{eq:Z00Result}
and \eqref{eq:Z0zResult}, respectively, and $\Sigma_{yy}=-\left(\Delta/i\omega\right)\Sigma_{00}$.
Substituting these results into Eq.~\eqref{eq:CmDef}, we transform
$\mathcal{C}_{\alpha\beta}^{\mathrm{m}}(\omega)$ to
\begin{align*}
&\mathcal{C}_{\alpha\beta}^{\mathrm{m}}(\omega)  =\frac{C_{h}h_{0}^{2}}{\pi}\nu\left\langle v_{\alpha}v_{\beta}\right\rangle \\
&\times  \frac{\partial}{\partial\omega}\frac{\omega}{\left(\omega^{2}\!+\!\Delta_{0}^{2}\right)^{3/2}}
\mathrm{Re}\!\!\int\limits _{-\infty}^{\infty}\!dz
\frac{\frac{-\omega^{2}+3\Delta_{0}^{2}}{\omega^{2}+\Delta_{0}^{2}}+z^{2}+2iz}{\left(z^{2}+1\right)^{2}}W\left(z,g\right),
\end{align*}
where $z\!=\!\xi/\sqrt{\omega^{2}\!+\!\Delta_{0}^{2}}$. The parameter $g\equiv g_{n}$
and the function $W\left(z,g\right)$ are defined in Eqs.~\eqref{eq:Def-gn}
and \eqref{eq:UResult}, respectively. Computation of the $z$ integral
yields the result 
\begin{widetext}
\begin{align}
\mathcal{C}_{\alpha\beta}^{\mathrm{m}}(\omega) & =-4C_{h}h_{0}^{2}\nu\left\langle v_{\alpha}v_{\beta}\right\rangle \frac{\partial}{\partial\omega}\left\{ \frac{\omega\Delta_{0}^{2}}{\left(\omega^{2}+\Delta_{0}^{2}\right)^{5/2}\left(4-g^{2}\right)}\left[1-\frac{6-g^{2}}{\sqrt{4-g^{2}}}\ln\left(\frac{2+\sqrt{4-g^{2}}}{g}\right)\right]\right\} .\label{eq:CmDeriv}
\end{align}
Therefore, the corresponding correction to the kernel, Eq.~\eqref{eq:KernCorrFull},
is 
\begin{equation}
\delta Q_{\alpha\beta}^{\mathrm{m}}=-2\frac{e^{2}}{c}C_{h}h_{0}^{2}\nu\left\langle v_{\alpha}v_{\beta}\right\rangle T\sum_{\omega_{n}}\frac{\partial}{\partial\omega_{n}}\left\{ \frac{\omega_{n}\Delta_{0}^{2}}{\left(\omega_{n}^{2}+\Delta_{0}^{2}\right)^{5/2}\left(4-g_{n}^{2}\right)}\left[1-\frac{6-g_{n}^{2}}{\sqrt{4-g_{n}^{2}}}\ln\left(\frac{2+\sqrt{4-g_{n}^{2}}}{g_{n}}\right)\right]\right\} .\label{eq:KernelCorrResult}
\end{equation}
\end{widetext}
This result gives correction at the fixed gap parameter. The full
correction also contains the contribution due to the shift of $\Delta$,
$\delta Q_{\alpha\beta}^{\Delta}=\tilde{\Delta}dQ_{\alpha\beta}^{(0)}/d\Delta$,
where $\tilde{\Delta}$ is given by Eq.~\eqref{eq:GapCorrRed}. Using
the same reduced variables as in Eq.~\eqref{eq:VD-T-a}, we
rewrite the corresponding correction to $\lambda_{\alpha}^{-2}\propto Q_{\alpha\alpha}$
in the reduced form suitable for numerical evaluation
\begin{subequations}
\begin{equation}
\lambda_{1\alpha}^{-2}(T)  =\!-\lambda_{0\alpha}^{-2}(T)\frac{h_{0}^{2}}{2\Delta_{0}^{2}\ln\left(\xi_{h}/a\right)}\mathcal{V}_{Q}\!\left(\frac{2\pi T}{\Delta_{0}},\frac{\xi_{s}}{\xi_{h}}\right)\label{eq:LambCorrCorrFl}
\end{equation}
with
\begin{align}
\mathcal{V}_{Q}\!\left(\tilde{T},\alpha_{h}\right)\! & =\left[\mathcal{D}(\tilde{T})\right]^{-1}\!\tilde{T}\sum_{n=0}^{\infty}\left[ K_{Q}\left(\tilde{\omega}_n\right)\mathcal{V}_{\Delta}\left(\tilde{T},\alpha_{h}\right)\right.\nonumber\\
&+\left.R_{Q}\left(\tilde{\omega}_n,\alpha_{h}\right)\right] ,\label{eq:VQ}\\
K_{Q}(z) & =-\frac{\partial}{\partial z}\frac{z}{\left(z^{2}\!+\!1\right)^{3/2}},\label{eq:KQ}\\
R_{Q}\left(z,\alpha_{h}\right) & =\frac{\partial}{\partial z}\frac{z\left[1-\left(3-\frac{2\alpha_{h}^{2}}{z^{2}\!+\!1}\right)L(z,\alpha_{h})\right]}{\left(z^{2}\!+\!1\right)^{3/2}\left(z^{2}\!+\!1\!-\!\alpha_{h}^{2}\right)},\label{eq:RQ}
\end{align}
\end{subequations}
where the first term in the square brackets in Eq.~\eqref{eq:VQ}
is due to the gap correction, the function $\mathcal{V}_{\Delta}(\tilde{T},\alpha_{h})$
is defined in Eq.~\eqref{eq:VD-T-a}, and the function $L(z,\alpha_{h})$
in the last definition is defined in Eq.~\eqref{eq:LogFun}. For brevity, in Eq.\ \eqref{eq:LambCorrCorrFl} we omitted the $T$ dependences of $\Delta_0(T)$, $\xi_s(T)$, and $\xi_h(T)$. Plots
of the function $\mathcal{V}_{Q}(\tilde{T},\alpha_{h})$
versus $\alpha_{h}$ for different values of $\tilde{T}$ are shown
in Fig.~\ref{Fig:VQ}. As the function $\mathcal{V}_{\Delta}(\tilde{T},\alpha_{h})$ shown in Fig.~\ref{Fig:VD-a-T}, this function also monotonically decreases with increasing of both $\tilde{T}$ and $\alpha_{h}$. The essential difference is that the function  $\mathcal{V}_{Q}(\tilde{T},\alpha_{h})$ vanishes for $\tilde{T}\!\rightarrow\!0$ while the function  $\mathcal{V}_{\Delta}(\tilde{T},\alpha_{h})$ approaches the finite limit.

The large-$\alpha_{h}$ asymptotics of the function $\mathcal{V}_{Q}(\tilde{T},\alpha_{h})$
is $\mathcal{V}_{Q}(\tilde{T},\alpha_{h})\approx\frac{\pi}{\alpha_{h}}V_{\lambda,m}(\tilde{T})$,
where the function $V_{\lambda,m}(\tilde{T})$ is defined
in Eq.~\eqref{eq:Vlm}. These asymptotics are also shown in Fig.~\ref{Fig:VQ}
by dashed lines. Noting also the relation $\pi h_{0}^{2}/\left(2\alpha_{h}\Delta_{0}\ln\left(\xi_{h}/a\right)\right)=1/\tau_{m}$,
we see that in the limit $\alpha_{h}\gg1$ the above result reproduces
the correction in Eq.~\eqref{eq:LambCorrScatt} for the scattering
regime. 

\begin{figure}[t]
\includegraphics[width=3.05in]{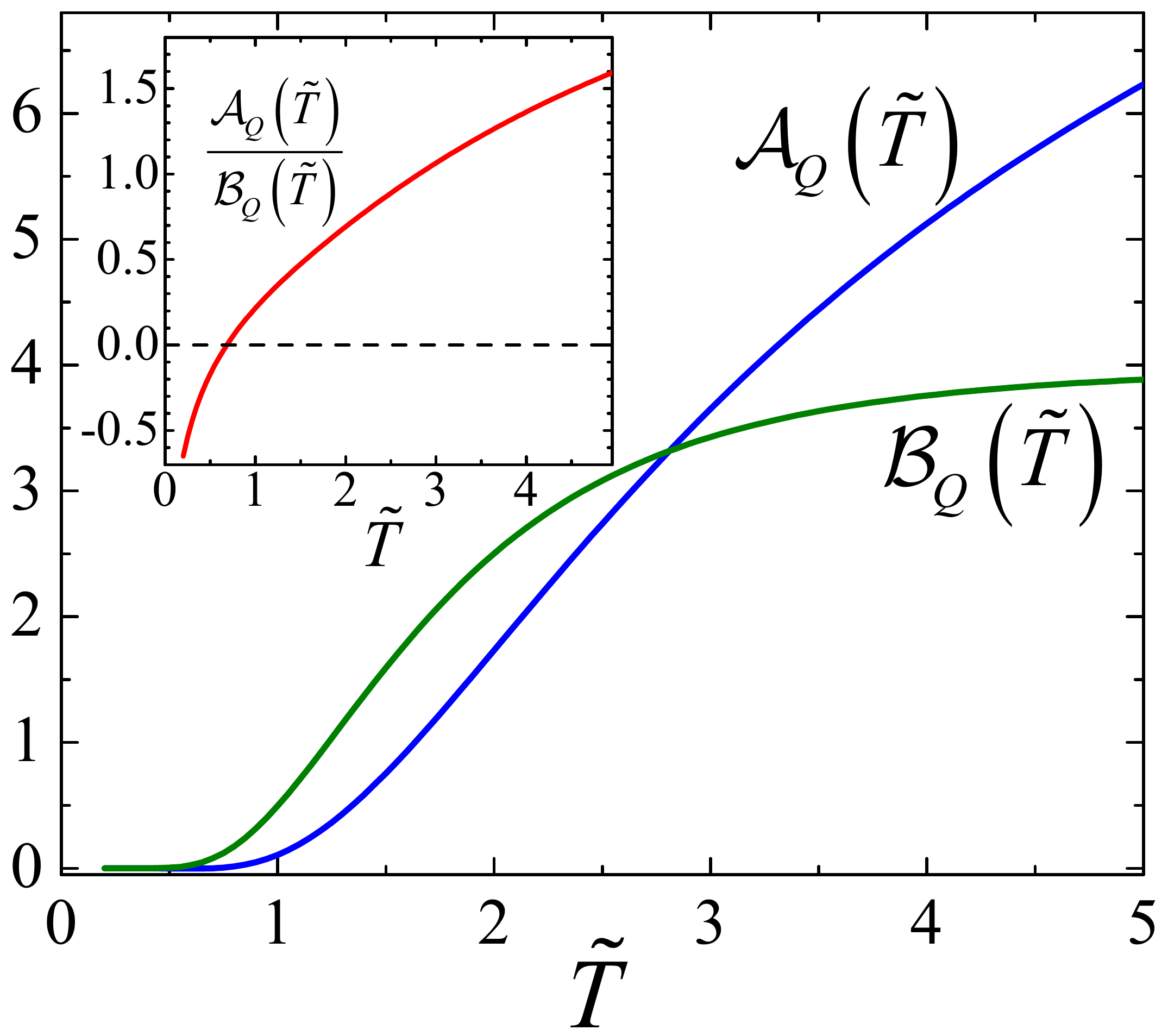}\caption{Temperature dependence of the coefficients $\mathcal{A}_{Q}(\tilde{T})$
and $\mathcal{B}_{Q}(\tilde{T})$ defined by Eqs.~\eqref{eq:AQ}
and \eqref{eq:BQ}, respectively, which determine the small-$\alpha_{h}$
asymptotics of the function $\mathcal{V}_{Q}(\tilde{T},\alpha_{h})$,
Eq.~\eqref{eq:VQ-Small-ah}. The inset shows the temperature dependence
of their ratio. The coefficient $\mathcal{A}_{Q}(\tilde{T})$ changes
sign at $\tilde{T}=0.683$.\label{fig:Lambda-small-ah} }
\end{figure}

At small $\alpha_{h}$ corresponding to the proximity of the magnetic
transition, the function $R_{Q}\left(z,\alpha_{h}\right)$ has logarithmic
dependence on $\alpha_{h}$,
\begin{align*}
R_{Q}\left(z,\alpha_{h}\right) & \approx R_{Q,0}(z)+R_{Q,1}(z)\ln\left(\frac{1}{\alpha_{h}}\right),\\
R_{Q,0}(z) & =\frac{\partial}{\partial z}\frac{z\left[1\!-\!3\ln\left(2\sqrt{z^{2}\!+\!1}\right)\right]}{\left(z^{2}\!+\!1\right)^{5/2}},\nonumber \\
R_{Q,1}(z) & =-\frac{\partial}{\partial z}\frac{3z}{\left(z^{2}\!+\!1\right)^{5/2}}.\nonumber 
\end{align*}
The function $\mathcal{V}_{\Delta}(\tilde{T},\alpha_{h})$
describing the gap contribution also has logarithmic dependence on
$\alpha_{h}$, Eq.~\eqref{eq:VDsmall-ah}. Correspondingly, the function
$\mathcal{V}_{Q}(\tilde{T},\alpha_{h})$ also logarithmically
diverges with $\alpha_{h}\rightarrow0$,
\begin{subequations}
	\begin{equation}
	\mathcal{V}_{Q}\!\left(\tilde{T},\alpha_{h}\right)  =
	\mathcal{A}_{Q}(\tilde{T})+\mathcal{B}_{Q}(\tilde{T})\ln\left(\frac{1}{\alpha_{h}}\right),
	\label{eq:VQ-Small-ah}
	\end{equation}
with
\begin{align}
&\mathcal{A}_{Q}(\tilde{T})\!= \!\left[\mathcal{D}(\tilde{T})\right]^{-1}\!\tilde{T}
\sum_{n=0}^{\infty}\!\left[ K_{Q}(\tilde{\omega}_n)\mathcal{A}(\tilde{T})\!+\!R_{Q,0}(\tilde{\omega}_n)\right], \label{eq:AQ}\\
&\mathcal{B}_{Q}(\tilde{T})\!= \!\left[\mathcal{D}(\tilde{T})\right]^{-1}\!\tilde{T}
\sum_{n=0}^{\infty}\!\left[ K_{Q}(\tilde{\omega}_n)\mathcal{B}(\tilde{T})\!+\!R_{Q,1}(\tilde{\omega}_n)
\right], 
\label{eq:BQ}
\end{align}
\end{subequations}
where the coefficients $\mathcal{A}(\tilde{T})$ and $\mathcal{B}(\tilde{T})$
are defined in Eqs.~\eqref{eq:AVDsmall-ah} and \eqref{eq:BVDsmall-ah}, respectively. The small-$\alpha_{h}$ asymptotics
are plotted in Fig.~\ref{Fig:VQ} with dotted lines and plots of the  
the coefficients $\mathcal{A}_{Q}(\tilde{T})$
and $\mathcal{B}_{Q}(\tilde{T})$ and their ratio are presented in Fig.~\ref{fig:Lambda-small-ah}. Note that the coefficient $\mathcal{A}_{Q}(\tilde{T})$
becomes negative for $\tilde{T}\!<\!0.683$.
Even though the small-$\alpha_{h}$ behavior in Eq.\ \eqref{eq:VQ-Small-ah} looks similar to the behavior of the gap in Eq.~\eqref{eq:VDsmall-ah}, the essential difference is that
both coefficients $\mathcal{A}_{Q}(\tilde{T})$ and $\mathcal{B}_{Q}(\tilde{T})$
vanish at $\tilde{T}=0$. 

Similar to Eq.~\eqref{eq:DeltaLarge-xih}, we can present the correction
in Eq.~\eqref{eq:LambCorrCorrFl} in the limit $\xi_{h}\gg\xi_{s}$
as
\begin{equation}
\lambda_{1\alpha}^{-2}\!=\!-\!\lambda_{0\alpha}^{-2}\frac{h_{0}^{2}\mathcal{B}_{Q}\!(\tilde{T})}{2\Delta_{0}^{2}}\!\left[1\!-\!\frac{\ln\left(\frac{\xi_{s}}{a}\right)\!-\!\mathcal{A}_{Q}\!(\tilde{T})\!/\!\mathcal{B}_{Q}\!(\tilde{T})}{\ln\left(\xi_{h}/a\right)}\right].\label{eq:LambLarge-xih}
\end{equation}
The ratio $\mathcal{A}_{Q}\!(\tilde{T})\!/\!\mathcal{B}_{Q}\!(\tilde{T})$
is of the order unity in the whole temperature range and becomes negative
for $\tilde{T}<0.683$, see inset in Fig.~\ref{fig:Lambda-small-ah},
meaning that the nominator $\ln\left(\xi_{s}/a\right)\!-\!\mathcal{A}_{Q}\!(\tilde{T})\!/\!\mathcal{B}_{Q}\!(\tilde{T})$
is always positive. As a consequence, the correction to the superfluid
density monotonically increases when temperature approaches $T_{m}$.
This is different from the behavior of the gap correction, Eq.~\eqref{eq:GapCorrRed},
which becomes nonmonotonic at small temperatures. The maximum suppression
of $\lambda_{\alpha}^{-2}$ for $\xi_{h}\rightarrow\infty$, $\lambda_{1\alpha,\mathrm{max}}^{-2}\!=\!-\lambda_{0\alpha}^{-2}\left(h_{0}^{2}/2\Delta_{0}^{2}\right)\mathcal{B}_{Q}\!\left(2\pi T_{m}/\Delta_{0}\right)$,
corresponds to the correction from a uniform exchange field equal
to $h_{0}$.

\section{Discussion\label{sec:Discussion}}
\begin{figure*}
	\includegraphics[width=3.2in]{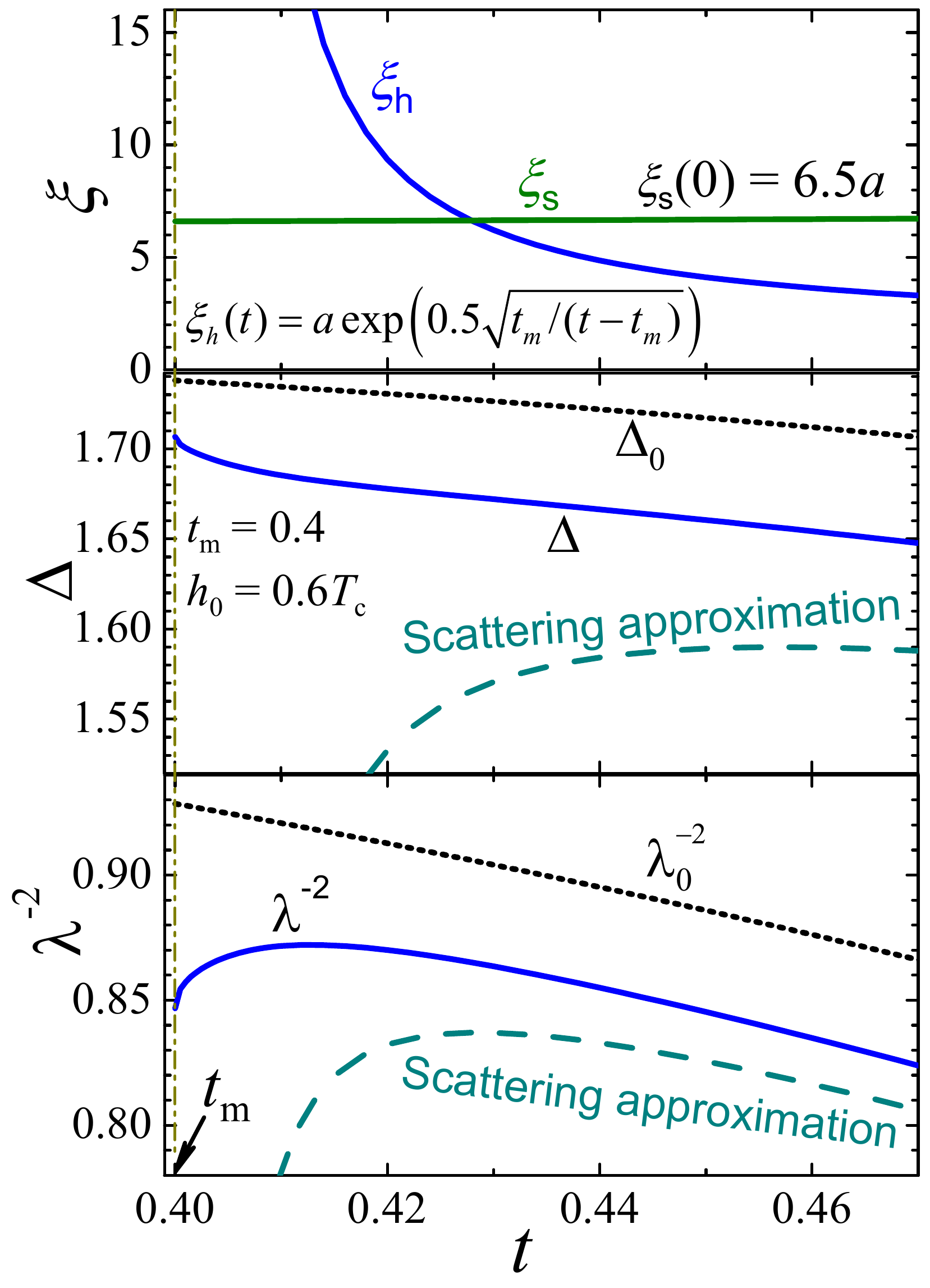}\includegraphics[width=3.15in]{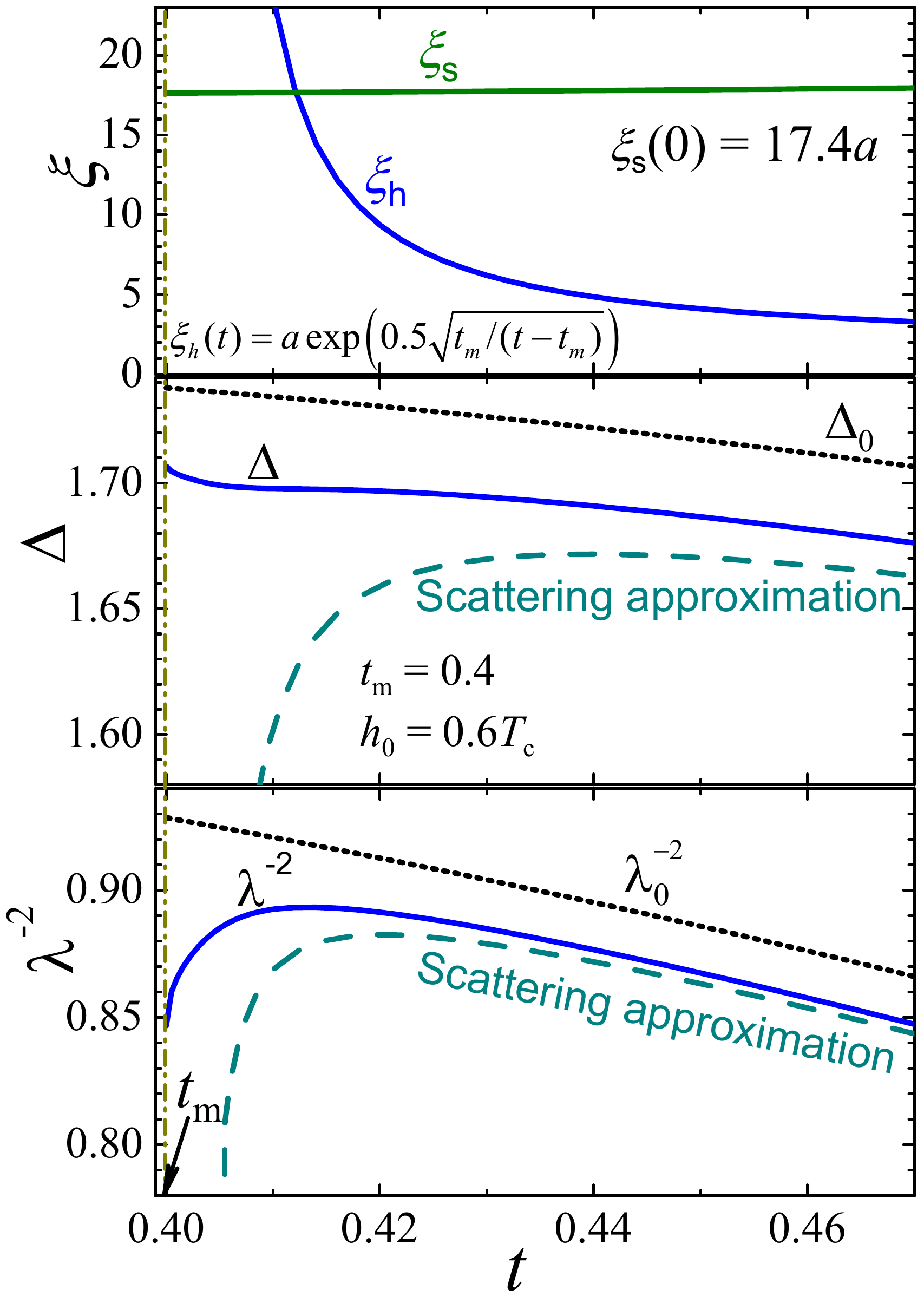}
	\caption{The middle and bottom panels in both plots show computed dependences of the gap $\Delta$ and superfluid density $\propto\lambda^{-2}$ on the reduced
		temperature $t=T/T_{c}$. The dotted lines show unperturbed values
		and dashed lines show the results obtained within the scattering approximation.
		The unit of $\Delta$ is $T_{c}$ and the unit of $\lambda^{-2}$ is $\left[\lambda_{0}(0)\right]^{-2}$. The top panel shows the assumed temperature dependences of the magnetic
		correlation length and coherence length.
		The plots on the left side are made for the parameters roughly corresponding
		to RbEuFe$_{4}$As$_{4}$ (see text). The plots on the right side are made  for the same parameters as in the left plots except for 2.7 times larger coherence length. In this case the scattering-regime asymptotics are much more pronounced and the gap has nonmonotonic temperature dependence. 		
		}
	\label{Fig:IllustrTdep}
\end{figure*}

In summary, we evaluated the corrections to the gap, Eq.~\eqref{eq:GapCorrRed},
and superfluid density, Eq.~\eqref{eq:LambCorrCorrFl}, caused by
the exchange interaction with quasi-two-dimensional magnetic fluctuations in  materials composed of superconducting and local-moment layers. Growth of the correlation length near the magnetic transition enhances spin-flip scattering leading to increasing suppression of superconducting parameters. This suppression significantly weakens when the magnetic correlation length exceeds the coherence length.
In addition to dependence on the correlation
length $\xi_{h}(T)$, the corrections have also direct
regular dependence on the ratio $T/\Delta_{0}(T)$.  
Moreover, as one can see from Figs.~\ref{Fig:VD-a-T}
and \ref{Fig:VQ}, in the paramagnetic state these dependences are
opposite. While in the immediate vicinity of the magnetic transition
the growth of $\xi_{h}(T)$ dominates, in a wider range, 
the overall temperature dependence is determined
by the interplay between both sources. To generate the parameter's
temperature dependences for real materials from the derived general formulas,
one need to specify the temperature dependent gap, coherence length,
and magnetic correlation length, as well as the strength of the exchange
field.

Even though the consideration of this paper has been mostly motivated by
physics of RbEuFe$_{4}$As$_{4}$, at present, there are too many uncertainties
in the parameters of this material to make a reliable quantitative
predictions. Therefore, we limit ourselves with showing expected qualitative
behavior using representative parameters and illustrating general
trends. Figure \ref{Fig:IllustrTdep}(left) shows the temperature dependences
of the gap and $\lambda^{-2}$ for the parameters very roughly corresponding
to RbEuFe$_{4}$As$_{4}$. Namely, we assume (i)the the Ginzburg-Landau
coherence length $\xi_{s0}^{\mathrm{GL}}\!=\!1.46$nm, following from
the linear slope of the c-axis upper critical field \cite{Smylie2018,WillaPhysRevB.99.180502},
(ii) the BCS value of the zero-temperature gap, $\Delta_{0}(0)\!=\!1.76T_{c}\!\approx\!5.6$meV,
(iii) the BCS temperature dependences for all unperturbed superconducting
parameters, (iv) the amplitude of the exchange field $h_{0}\!=\!0.6T_{c}$,
(v) the magnetic transition at $t_{m}\!\equiv\!T_{m}/T_{c}\!=\!0.4$,
and (vi) Berezinskii-Kosterlitz-Thouless (BKT) shape for the magnetic correlation
length, $\xi_{h}(T)\!=\!a\exp[b\sqrt{T_{m}/(T\!-\!T_{m})}]$, where $a\!=\!0.39$
nm is the distance between the neighboring Eu$^{2+}$ moments and
we take the value $b\!=\!0.5$ for nonuniversal numerical constant.
For these parameters, $\xi_{s}(T_{m})=6.6a$ and the 'scattering-to-smooth' crossover is nominally located at $t_{\mathrm{cr}}\!\approx\!0.43$.
We see, however, that above this temperature the behavior is not well
described by the scattering-regime asymptotics shown by the dashed
lines. This is related to the broad range of the crossover. Consequently,
for selected parameters, the gap does not display a nonmonotonic behavior,
expected from the analysis of asymptotics. In fact, due to the interplay
between two competing temperature dependences, both corrections are
almost temperature independent in the range $0.42<t<0.47$. Nevertheless,
we see that, according to the general predictions, $\Delta(T)$ somewhat
increases when $T$ approaches $T_{m}$, while $[\lambda(T)]^{-2}$
shows a noticeable drop. For illustrative purposes, we show in Fig.~\ref{Fig:IllustrTdep}(right)
the plots of $\Delta(T)$ and $[\lambda(T)]^{-2}$ for the same parameters
as in the previous figure except for larger coherence length, $\xi_{s0}^{\mathrm{GL}}\!=\!10a\!\approx\!3.9$nm.
In this case $\xi_{s}(T_{m})\!=\!17.6a$ and the crossover nominally
takes place much closer to $t_{m}$, at $t_{\mathrm{cr}}\!\approx\!0.41$.
In this case the behavior at $t>0.43-0.44$ is already fairly well
described by the scattering asymptotics. The gap in this case does
have a nonmonotonic temperature dependence.

Clearly, the plots in Fig.~\ref{Fig:IllustrTdep}(left) do not literally
describe the behavior of RbEuFe$_{4}$As$_{4}$ and serve only as
a qualitative illustration. This material has several additional features
that influence the behavior of the parameters but substantially complicate
an accurate analysis. Firstly, the assumed two-dimensional behavior always
breaks down sufficiently close to the transition and the dimensional
crossover to the three-dimensional regime takes place. In this 3D regime
the correlations between the different magnetic layers emerge meaning
that the assumption for two-dimensional scattering does not work any
more. In addition, the magnetic correlation length does not follow
the BKT temperature dependence assumed in Fig.~\ref{Fig:IllustrTdep}.
Secondly, due to spatial separation between the magnetic and conducting
layers, we expect a significant nonlocality of the exchange interaction,
see Eq.~\eqref{eq:ExField}, ranging at least 2--3 lattice spacing.
Consideration of this manuscript assumes that the magnetic correlation
length exceeds this nonlocality range. This assumption is only justified
close to the magnetic transition. The nonlocality significantly reduces
the exchange corrections at higher temperatures, when $\xi_{h}$ drops
below the nonlocality range. Finally, our single-band consideration does not take into account a complicated multiple-band structure of RbEuFe$_{4}$As$_{4}$. 

In this paper, we developed a general theoretical framework for the analysis of the influence of correlated magnetic fluctuations on superconducting parameters.
We focus on the behavior of the gap and superfluid density for the in-plane current direction, but the consideration can be directly extended to other thermodynamic and transport properties. 
For some properties, however,  such as specific heat and magnetization, a reliable separation of the superconducting contribution from the magnetic background in experiment is challenging. This makes a theoretical analysis somewhat academic. 
Our result can be straightforwardly generalized to the case of a large exchange field leading to a strong suppression of superconductivity. Such generalization requires the development of a self-consistent scheme similar to the AG theory\cite{AbrGorJETP61}. For the problem considered here, this is a formidable theoretical task.  

\begin{acknowledgments}
I would like to thank U.\ Welp, S.\ Bending, D.\ Collomb, and V. Kogan for useful discussions. This work was supported by the US Department of Energy, Office of
Science, Basic Energy Sciences, Materials Sciences and Engineering Division.
\end{acknowledgments}

\appendix
\section{Calculation of the integral for the gap correction \label{app:CalcInt}}

In this appendix we briefly describe calculation of the integral in Eq.\ \eqref{eq:CorrGapInt} leading to result in Eq.\ \eqref{eq:CorrGapIntResult}. Substituting the function $W(z,g)$ defined in Eq.~\eqref{eq:UResult}
into Eq.~\eqref{eq:CorrGapInt}, we present $\delta\mathcal{I}$ as
\begin{equation}
\delta\mathcal{I}\!= \!-\frac{C_{h}\nu h_{0}^{2}}{4\pi}
\frac{\Delta_{0}}{\left(\omega_{n}^{2}\!+\!\Delta_{0}^{2}\right)^{3/2}}
\left[\mathcal{T}_{1}(g_{n})\!+\!\frac{4\omega_{n}^{2}}{\omega_{n}^{2}\!+\!\Delta_{0}^{2}}\mathcal{T}_{2}(g_{n})\right],
\label{eq:dI-split}
\end{equation}
where 
\begin{align*}
\mathcal{T}_{1}(g) & =\mathrm{Re}
\left[\int\limits_{-\infty}^{\infty}dz\frac{2}{\left(z+i\right)^{2}\sqrt{\left(iz+1\right)^{2}-g^{2}}}\right.\\
\times&\left.\ln\left(\frac{iz+1+\sqrt{\left(iz+1\right)^{2}-g^{2}}}{g}\right)\right],\\
\mathcal{T}_{2}(g) & =\mathrm{Re}
\left[\int\limits_{-\infty}^{\infty}dz\frac{2}{\left(z^{2}+1\right)^{2}\sqrt{\left(iz+1\right)^{2}-g^{2}}}\right.\\
\times&\left.\ln\left(\frac{iz+1+\sqrt{\left(iz+1\right)^{2}-g^{2}}}{g}\right)\right].
\end{align*}
The integral for $\mathcal{T}_{1}(g)$ has a pole at
$z=-i$ and branches at the imaginary axis terminating at $z_{\pm}=i\left(1\pm g\right)$.
Deforming the integration contour into the complex plane, we reduce
it to the integral along the square-root branch $z=ix$, $1+g<x<\infty$
, 
\begin{align*}
\mathcal{T}_{1}(g) & =-2\pi\int\limits _{1+g}^{\infty}dx\frac{2}{\left(x+1\right)^{2}\sqrt{\left(x-1\right)^{2}-g^{2}}}\\
 & =4\pi\left[\frac{1}{4-g^{2}}-\frac{2}{\left(4-g^{2}\right)^{3/2}}\ln\left(\frac{2+\sqrt{4-g^{2}}}{g}\right)\right].
\end{align*}
The integral for $\mathcal{T}_{1}(g)$ has the same square-root branches
and the poles at $z=\pm i$. Consequently, we split the integral into
contribution from the pole at $z=i$ and square-root branch $z=ix$,
$1+g<x<\infty$ which yields
\begin{align*}
\mathcal{T}_{2}(g) & =\frac{\pi^{2}}{2g}-\frac{\pi}{g^{2}}+4\pi\int\limits _{1+g}^{\infty}dx\frac{1}{\left(x^{2}-1\right)^{2}\sqrt{\left(x-1\right)^{2}-g^{2}}}\\
 & =-\pi\left[\frac{1}{4-g^{2}}-\frac{6-g^{2}}{\left(4-g^{2}\right)^{3/2}}\ln\left(\frac{2+\sqrt{4-g^{2}}}{g}\right)\right]
\end{align*}
Substituting the above results into Eq.~\eqref{eq:dI-split}, we
arrive to Eq.~\eqref{eq:CorrGapIntResult}.

\section{Magnetic-scattering correction to the London penetration depth using
quasiclassical approach\label{app:Magn-Scatt-Lond} }

The London penetration depth $\lambda$ in the presence of isotropic
potential and magnetic scattering has been investigated within the quasiclassical
approach in Ref.~\cite{KoganLonPenDepthPhysRevB13}. Here we derive
a general equation for $\lambda$ for arbitrary magnetic scattering
having in mind application to the case of correlated magnetic fluctuations.
The Eilenberger equations for the quasiclassical Green's functions, $f(\boldsymbol{p},\boldsymbol{r})$,
$f^{\dagger}(\boldsymbol{p},\boldsymbol{r})$, and $g(\boldsymbol{p},\boldsymbol{r})$
for arbitrary scattering are \cite{EilenbergerZPhys1968} 
\begin{subequations}
\begin{align}
 & \bm{v}\bm{\Pi}f=2\Delta g-2\omega_{n}f\nonumber \\
 & +g\left\langle \left[W(\boldsymbol{p},\boldsymbol{p}^{\prime})\!-W_{m}(\boldsymbol{p},\boldsymbol{p}^{\prime})\right]f{}^{\prime}\right\rangle ^{\prime}\!\nonumber \\
 & -\!f\left\langle \left[W(\boldsymbol{p},\boldsymbol{p}^{\prime})\!+\!W_{m}(\boldsymbol{p},\boldsymbol{p}^{\prime})\right]g{}^{\prime}\right\rangle ^{\prime},\label{EilArbSc1}\\
-\! & \!\bm{v}\bm{\Pi}^{*}\!f^{\dagger}\!=\!2\Delta^{*}g\!-\!2\omega_{n}f^{\dagger}\!\nonumber \\
 & +\!g\left\langle \left[W(\boldsymbol{p},\boldsymbol{p}^{\prime})\!-\!W_{m}(\boldsymbol{p},\boldsymbol{p}^{\prime})\right]f^{\dagger}{}^{\prime}\right\rangle ^{\prime}\!\nonumber \\
 & -\!f^{\dagger}\left\langle \left[W(\boldsymbol{p},\boldsymbol{p}^{\prime})\!+\!W_{m}(\boldsymbol{p},\boldsymbol{p}^{\prime})\right]g{}^{\prime}\right\rangle ^{\prime}\!,\label{EilArbSc2-1}
\end{align}
\end{subequations}where we used shortened notations $f\!\equiv\!f(\boldsymbol{p},\boldsymbol{r})$,
$f^{\prime}\!\equiv\!f(\boldsymbol{p}^{\prime},\boldsymbol{r})$,
$\bm{\Pi}f\!\equiv\!\left(\boldsymbol{\nabla}\!+\!2\pi i\bm{\bm{A}}/\phi_{0}\right)f$,
$\left\langle A(\boldsymbol{p}^{\prime})\right\rangle ^{\prime}\!\equiv\!\int_{S_{F}}\!d^{2}\boldsymbol{p}^{\prime}\rho(\boldsymbol{p}^{\prime})A(\boldsymbol{p}^{\prime})$,
and $\rho(\boldsymbol{p})\!=\!\left[(2\pi)^{3}\nu v_{F}(\boldsymbol{p})\right]^{-1}$.
Further, $W(\boldsymbol{p},\boldsymbol{p}^{\prime})$ and $W_{m}(\boldsymbol{p},\boldsymbol{p}^{\prime})$
are the probabilities of potential and magnetic scattering defining
the corresponding scattering times as
\[
\frac{1}{\tau}=\left\langle W(\boldsymbol{p},\boldsymbol{p}^{\prime})\right\rangle ^{\prime},\,\frac{1}{\tau_{m}}=\left\langle W_{m}(\boldsymbol{p},\boldsymbol{p}^{\prime})\right\rangle ^{\prime}.
\]
For the model considered in this paper $W_{m}(\boldsymbol{p},\boldsymbol{p}^{\prime})=2\pi\nu\left\langle \left|\tilde{\boldsymbol{h}}_{\boldsymbol{p}-\boldsymbol{p}^{\prime}}\right|^{2}\right\rangle $
. The above equations have to be supplemented with the normalization
condition $g^{2}=1-ff^{\dagger}$, the gap equation
\begin{equation}
\frac{\Delta}{2\pi T}\ln\frac{T_{c0}}{T}=\sum_{\omega_{n}>0}\left(\frac{\Delta}{\omega}-\langle f\rangle\right),\label{EilArbScGap}
\end{equation}
 and formula for the current
\begin{equation}
\bm{j}=4\pi e\nu T\,{\rm Im}\sum_{\omega_{n}>0}\langle\bm{v}g\rangle.\label{EilArbScCurr}
\end{equation}

Our goal is to derive the response to weak supercurrents. In linear
order, weak supercurrents do not modify the gap absolute value but
only add the same phase $\theta(\bm{r})$ to $\Delta$ and $f$ and
the opposite phase to $f^{\dagger}$. Therefore, the linear-order
solutions have the form, 
\begin{align*}
\Delta=\Delta_{0}\,e^{i\theta},\,f=(f_{0}+f_{1})\,e^{i\theta},\\
f^{\dagger}=(f_{0}+f_{1})e^{-i\theta},\,g=g_{0}+g_{1}\,,
\end{align*}
where only the phase $\theta$ has coordinate dependence, while $f_{1},f_{1}^{\dagger},g_{1}$
are uniform meaning that $\bm{\Pi}f=i\bm{P}f$ and $\bm{\Pi}^{*}f^{\dagger}=-i\bm{P}f^{\dagger}$
with $\bm{P}=\nabla\theta+2\pi\bm{\bm{A}}/\phi_{0}$. Equations for
the linear corrections are
\begin{align}
2\Delta_{0}&g_{1}\!  -\!2\omega_{n}f_{1}\!+\!\frac{g_{1}(\boldsymbol{p})}{\tau_{-}}f_{0}-\frac{f_{1}(\boldsymbol{p})}{\tau_{+}}g_{0}\nonumber \\
+ & g_{0}\left\langle \left[W(\boldsymbol{p},\boldsymbol{p}^{\prime})\!-\!W_{m}(\boldsymbol{p},\boldsymbol{p}^{\prime})\right]f_{1}(\boldsymbol{p}^{\prime})\right\rangle ^{\prime}\!\nonumber \\
 -&f_{0}\left\langle \left[W(\boldsymbol{p},\boldsymbol{p}^{\prime})\!+\!W_{m}(\boldsymbol{p},\boldsymbol{p}^{\prime})\right]g_{1}(\boldsymbol{p}^{\prime})\right\rangle ^{\prime}\!=\!iv_{\alpha}P_{\alpha}f_{0},\label{EilArbScLin1}\\
 & g_{0}g_{1}=-f_{0}f_{1}\label{EilArbScLin2}
\end{align}
with $\frac{1}{\tau_{\pm}}=\frac{1}{\tau}\pm\frac{1}{\tau_{m}}.$
The remaining averages in the first equation account for the reverse
scattering events. These averages vanish for the case of isotropic
scattering. We assume that the solutions are proportional to $v_{\alpha}P_{\alpha}$
and define the corresponding averages as
\[
\left\langle W(\boldsymbol{p},\boldsymbol{p}^{\prime})v_{\alpha}^{\prime}\right\rangle ^{\prime}=\frac{1}{\tau^{\alpha}}v_{\alpha},\,\left\langle W_{m}(\boldsymbol{p},\boldsymbol{p}^{\prime})v_{\alpha}^{\prime}\right\rangle ^{\prime}=\frac{1}{\tau_{m}^{\alpha}}v_{\alpha},
\]
giving
\begin{equation}
\frac{1}{\tau^{\alpha}}=\frac{\left\langle \left\langle W(\boldsymbol{p},\boldsymbol{p}^{\prime})v_{\alpha}v_{\alpha}^{\prime}\right\rangle \right\rangle ^{\prime}}{\left\langle v_{\alpha}^{2}\right\rangle },\,\frac{1}{\tau_{m}^{\alpha}}=\frac{\left\langle \left\langle W_{m}(\boldsymbol{p},\boldsymbol{p}^{\prime})v_{\alpha}v_{\alpha}^{\prime}\right\rangle \right\rangle ^{\prime}}{\left\langle v_{\alpha}^{2}\right\rangle }.\label{eq:tau-alpha}
\end{equation}
These quantities determine the corresponding transport times in a
standard way, $1/\tau^{\mathrm{tr}}=1/\tau-1/\tau^{\alpha}$ and $1/\tau_{m}^{\mathrm{tr}}=1/\tau_{m}-1/\tau_{m}^{\alpha}$.
In the case of correlated magnetic fluctuation which we consider in
this paper, the transport rate is much smaller than the total scattering
rate. The averagings in Eq.~\eqref{EilArbScLin1} can now be performed
as, 
\[
\left\langle \left[W(\boldsymbol{p},\boldsymbol{p}^{\prime})\!\pm W_{m}(\boldsymbol{p},\boldsymbol{p}^{\prime})\right]f_{1}(\boldsymbol{p}^{\prime})\right\rangle ^{\prime}\!=\frac{f_{1}(\boldsymbol{p})}{\tau_{\pm}^{\alpha}}.
\]
with $1/\tau_{\pm}^{\alpha}=1/\tau^{\alpha}\pm1/\tau_{m}^{\alpha}$.
This allows us to rewrite Eq.~\eqref{EilArbScLin1} as
\begin{equation}
2\Delta_{0}g_{1}\!-\!2\omega_{n}f_{1}\!+\frac{g_{1}}{\tau_{-}}f_{0}\!+g_{0}\frac{f_{1}}{\tau_{-}^{\alpha}}\!-\frac{f_{1}}{\tau_{+}}g_{0}-f_{0}\frac{g_{1}}{\tau_{+}^{\alpha}}=iv_{\alpha}P_{\alpha}f_{0},\label{eq:EilArbScLin1-f}
\end{equation}
Substituting $g_{1}=-f_{0}f_{1}/g_{0}$ from Eq.~\eqref{EilArbScLin2},
we obtain the solutions
\begin{align}
f_{1} & =-\frac{\!iv_{\alpha}f_{0}P_{\alpha}}{2\omega_{n}\!+\!2\Delta_{0}\frac{f_{0}}{g_{0}}\!+\!\frac{f_{0}^{2}}{g_{0}}\left(\frac{1}{\tau_{-}}\!-\!\frac{1}{\tau_{+}^{\alpha}}\right)\!+\!g_{0}\left(\frac{1}{\tau_{+}}\!-\!\frac{1}{\tau_{-}^{\alpha}}\right)},\label{eq:Sol-f1}\\
g_{1} & =\frac{f_{0}^{2}}{g_{0}}\frac{\!iv_{\alpha}P_{\alpha}}{2\omega_{n}\!+\!2\Delta_{0}\frac{f_{0}}{g_{0}}\!+\!\frac{f_{0}^{2}}{g_{0}}\left(\frac{1}{\tau_{-}}\!-\!\frac{1}{\tau_{+}^{\alpha}}\right)\!+\!g_{0}\left(\frac{1}{\tau_{+}}\!-\!\frac{1}{\tau_{-}^{\alpha}}\right)}.
\label{eq:Sol-g1}
\end{align}
We can rewrite the scattering-rate differences here in terms of scattering
and transport times as
\begin{align*}
\frac{1}{\tau_{-}}\!-\!\frac{1}{\tau_{+}^{\alpha}} & =\frac{1}{\tau^{\mathrm{tr}}}-\frac{2}{\tau_{m}}+\frac{1}{\tau_{m}^{\mathrm{tr}}},\\
\frac{1}{\tau_{+}}\!-\!\frac{1}{\tau_{-}^{\alpha}} & =\frac{1}{\tau^{\mathrm{tr}}}+\frac{2}{\tau_{m}}-\frac{1}{\tau_{m}^{\mathrm{tr}}}.
\end{align*}

We use a standard parametrization for the unperturbed Green's function components 
\[
f_{0}=\frac{1}{\sqrt{1+u^{2}}},\:g_{0}=\frac{u}{\sqrt{1+u^{2}}}.
\]
The parameter $u$ obeys the Abrikosov-Gor'kov equation\cite{AbrGorJETP61,MakiInParks1969}
\begin{equation}
u\left(1-\frac{1}{\tau_{m}\Delta\sqrt{1+u^{2}}}\right)=\frac{\omega_{n}}{\Delta}.\label{eq:AGEq}
\end{equation}
 and determines the gap via equation
\begin{equation}
\Delta\ln\frac{T_{c0}}{T}=2\pi T\sum_{\omega_{n}>0}\left(\frac{\Delta}{\omega_{n}}-\frac{1}{\sqrt{1+u^{2}}}\right).\label{eq:Delta-u-par}
\end{equation}

We proceed with the derivation of the current response using  Eq.~\eqref{EilArbScCurr}. Rewriting $g_{1}$ in Eq.\ \eqref{eq:Sol-g1} in terms of the parameter $u$ 
\[
g_{1}  =\frac{\!iv_{\alpha}P_{\alpha}}{\left(1+u^{2}\right)\left(2\Delta\sqrt{1+u^{2}}+\!\frac{1}{\tau_{-}^{\mathrm{tr}}}\right)\!-\!\frac{2}{\tau_{m}^{\alpha}}}.
\]
and substituting it into Eq.~\eqref{EilArbScCurr}, we obtain the linear current response as
\begin{equation}
j_{\alpha} \!=\!4\pi e\nu T\sum_{\omega_{n}>0}\frac{\langle v_{\alpha}^{2}\rangle}{\left(1\!+\!u^{2}\right)\left(2\Delta\sqrt{1\!+\!u^{2}}+\!\frac{1}{\tau_{-}^{\mathrm{tr}}}\right)\!-\!\frac{2}{\tau_{m}^{\alpha}}}P_{\alpha}.\label{eq:CurrRespGen}
\end{equation}
Using the definition $4\pi j_{\alpha}/c=-\lambda_{\alpha}^{-2}A_{\alpha}$
and $P_{\alpha}=2\pi\,A_{\alpha}/\phi_{0}$, we finally obtain the
result for $\lambda_{\alpha}^{-2}$, 
\begin{equation}
\lambda_{\alpha}^{-2}\!=\!\frac{16\pi^{3}|e|\nu\langle v_{\alpha}^{2}\rangle}{c\phi_{0}}T\!\!\sum_{\omega_{n}\!>0}\frac{1}{\left(1\!+\!u^{2}\right)\left(\Delta\sqrt{1\!+\!u^{2}}+\!\frac{1}{2\tau_{-}^{\mathrm{tr}}}\right)\!-\!\frac{1}{\tau_{m}^{\alpha}}}.\label{eq:LambArbScat}
\end{equation}
This result can be used for self-consistent evaluation of the London
penetration depth of arbitrary scattering. Note that it is different
from the similar result in Ref.~\cite{SkalskiPhysRev.136.A1500}
by the sign in front of $\frac{1}{\tau_{m}^{\alpha}}$ in the denominator.

\subsubsection{Small-scattering-rate expansion}

For comparison with the results in the main text, we derive small
correction to $\lambda_{\alpha}^{-2}$ in the case of weak scattering.
Expanding the parameter $u$ in Eq.~\eqref{eq:AGEq}, $u=\frac{\omega_{n}}{\Delta}+u_{m},$
we obtain 
\[
u_{m}\approx\frac{\omega_{n}}{\tau_{m}\Delta\sqrt{\omega_{n}^{2}+\Delta^{2}}}.
\]
Substituting this expansion into the gap equation, Eq.~\eqref{eq:Delta-u-par},
we derive the correction to the gap, $\Delta=\Delta_{0}+\tilde{\Delta}$,
\begin{align}
\tilde{\Delta} & \!=\!-\frac{2\pi T}{\tau_{m}\Delta_{0}}\!\sum_{\omega_{n}>0}\frac{\omega_{n}^{2}}{\left(\Delta_{0}^{2}\!+\!\omega_{n}^{2}\right)^{2}}\left[2\pi T\!\!\sum_{\omega_{n}>0}\frac{1}{\left(\Delta_{0}^{2}\!+\!\omega^{2}\right)^{3/2}}\right]^{-1}\!.\label{eq:GapCorrQC}
\end{align}
In the reduced form, this correction is identical to Eq.~\eqref{eq:GapCorrMagnScat}. 

To derive the correction to $\lambda_{\alpha}^{-2}$, we expand the
fraction in Eq.~\eqref{eq:LambArbScat}
\begin{align*}
 & \frac{1}{\left(1\!+\!u^{2}\right)\left(\Delta\sqrt{1\!+\!u^{2}}\!+\!\frac{1}{2\tau_{-}^{\mathrm{tr}}}\right)\!-\!\frac{1}{\tau_{m}^{\alpha}}}\!\approx\!\frac{\Delta^{2}}{\left(\Delta^{2}\!+\!\omega_{n}^{2}\right)^{3/2}}\\
 & \times\left(\!1\!-\frac{3\omega_{n}^{2}-\Delta^{2}}{\tau_{m}\left(\Delta^{2}\!+\!\omega_{n}^{2}\right)^{3/2}}\!+\frac{\omega_{n}^{2}-\Delta^{2}}{2\tau_{m}^{\mathrm{tr}}\left(\Delta^{2}\!+\!\omega_{n}^{2}\right)^{3/2}}\!\right.\\
 & -\left.\frac{1}{2\tau^{\mathrm{tr}}\sqrt{\Delta^{2}\!+\!\omega_{n}^{2}}}\!\right)
\end{align*}
and also separate the contribution from the gap correction
\begin{align*}
\frac{\Delta^{2}}{\left(\Delta^{2}\!+\!\omega_{n}^{2}\right)^{3/2}} & \approx\frac{\Delta_{0}^{2}}{\left(\Delta_{0}^{2}+\omega_{n}^{2}\right)^{3/2}}+\frac{\Delta_{0}\tilde{\Delta}\left(2\omega_{n}^{2}\!-\!\Delta_{0}^{2}\right)}{\left(\Delta_{0}^{2}+\omega_{n}^{2}\right)^{5/2}}.
\end{align*}
This gives the correction to $\lambda_{\alpha}^{-2}$,
\begin{align}
\lambda_{\alpha}^{-2} & \approx\lambda_{0\alpha}^{-2}+\lambda_{1\alpha}^{-2},\nonumber \\
\lambda_{0\alpha}^{-2} & =\!\frac{16\pi^{4}\nu\langle v_{\alpha}^{2}\rangle}{\phi_{0}^{2}}T\sum_{\omega_{n}>0}\frac{\Delta_{0}^{2}}{\left(\Delta_{0}^{2}+\omega_{n}^{2}\right)^{3/2}},\nonumber \\
\lambda_{1\alpha}^{-2} & =\!\frac{16\pi^{4}\nu\langle v_{\alpha}^{2}\rangle}{\phi_{0}^{2}}T\!\sum_{\omega_{n}>0}\!\left[\frac{\Delta_{0}\tilde{\Delta}\left(2\omega_{n}^{2}\!-\!\Delta_{0}^{2}\right)}{\left(\Delta_{0}^{2}+\omega_{n}^{2}\right)^{5/2}}\right.\nonumber \\
- & \frac{\Delta_{0}^{2}\left(3\omega_{n}^{2}\!-\!\Delta_{0}^{2}\right)}{\tau_{m}\left(\Delta_{0}^{2}+\omega_{n}^{2}\right)^{3}}+\frac{\Delta_{0}^{2}\left(\omega_{n}^{2}\!-\!\Delta_{0}^{2}\right)}{2\tau_{m}^{\mathrm{tr}}\left(\Delta_{0}^{2}\!+\!\omega_{n}^{2}\right)^{3}}\nonumber \\
- & \left.\frac{\Delta_{0}^{2}}{2\tau^{\mathrm{tr}}\left(\Delta_{0}^{2}\!+\!\omega_{n}^{2}\right)^{2}}\right].\label{eq:LambdaCorr-1}
\end{align}
We see that, in contrast to the potential scattering, which only influences
the London penetration depth via the transport time, the magnetic-scattering
contribution to $\lambda_{1\alpha}^{-2}$ also has contributions proportional
to the total scattering rate $1/\tau_{m}$, both direct and via the gap
correction. This pair-breaking contribution, however, vanishes at
zero temperature. For numerical convenience, we can rewrite the correction
in the following reduced form
\begin{align}
&\lambda_{1\alpha}^{-2}  (T)=-\lambda_{0\alpha}^{-2}(T)\left[\frac{1}{\tau_{m}\Delta_{0}}
V_{\lambda,m}\!\left(\frac{2\pi T}{\Delta_{0}}\right)\nonumber \right.\\
&+\left.\frac{1}{\tau_{m}^{\mathrm{tr}}\Delta_{0}}
V_{\lambda,m}^{\mathrm{tr}}\!\left(\frac{2\pi T}{\Delta_{0}}\right)
\!+\!\frac{1}{\tau^{\mathrm{tr}}\Delta_{0}}
V_{\lambda}^{\mathrm{tr}}\!\left(\frac{2\pi T}{\Delta_{0}}\right)\right],
\label{eq:LambCorrArbScatt}
\end{align}
with
\begin{align}
V_{\lambda,m}(\tilde{T}) & \!=\!\left[\mathcal{D}\!(\tilde{T})\right]^{-1}\!\tilde{T}
\sum_{n=0}^{\infty}\!\left[\frac{2\tilde{\omega}_n^2\!-1}{\left(1\!+\!\tilde{\omega}_n^2\right)^{5/2}}V_{\Delta}(\tilde{T})\!+\!\frac{3\tilde{\omega}_n^2\!-1}{\left(1\!+\!\tilde{\omega}_n^2\right)^{3}}\right]\!,\nonumber \\
V_{\lambda,m}^{\mathrm{tr}}(\tilde{T}) & \!=\!\frac{1}{2}\left[\mathcal{D}(\tilde{T})\right]^{-1}\!\tilde{T}\sum_{n=0}^{\infty}\frac{1-\tilde{\omega}_n^2}{\left(1+\tilde{\omega}_n^2\right)^{3}},\nonumber \\
V_{\lambda}^{\mathrm{tr}}(\tilde{T}) & \!=\!\frac{1}{2}\left[\mathcal{D}(\tilde{T})\right]^{-1}\!\tilde{T}\sum_{n=0}^{\infty}\frac{1}{\left(1+\tilde{\omega}_n^2\right)^{2}},\nonumber 
\end{align}
where $\tilde{\omega}_n\!\equiv\!\tilde{T}(n\!+\tfrac{1}{2})$, $\mathcal{D}(\tilde{T})$ is defined in Eq.~\eqref{eq:DenomT}, and $V_{\Delta}(\tilde{T})$
in the formula for $V_{\lambda,m}(\tilde{T})$ is defined
in Eq.~\eqref{eq:VDScatt}. For zero temperature, we derive the following
result for the gap correction
\begin{align}
\lambda_{1\alpha}^{-2}(0) & =-\frac{\pi}{8}\lambda_{0\alpha}^{-2}\left(\frac{1}{2\tau_{m}^{\mathrm{tr}}\Delta_{0}}+\frac{1}{\tau^{\mathrm{tr}}\Delta_{0}}\right)\label{eq:LambCorrArbScatT0}
\end{align}
with $\lambda_{0\alpha}^{-2}=8\pi^{3}\nu\langle v_{\alpha}^{2}\rangle/\phi_{0}^{2}$.
Therefore, the correction to the London penetration depth at $T=0$
in the clean case is proportional to transport scattering rates for
both scattering channels.
\bibliography{CorrMagnFluct}

\end{document}